\documentclass[twocolumn,showpacs,amssymb,nobibnotes,aps,floats,psfig,prb]{revtex4-1}
\usepackage{textcomp,amssymb,graphicx,epsfig}
\usepackage{hyperref}
\usepackage{amsbsy}
\usepackage{amsmath,amsfonts,color}
\usepackage{IEEEtrantools}
\usepackage{float}
\date{}

\begin{document}
\title{A treatment of cooperative Jahn-Teller effect in interacting chains}
\author{Ravindra Pankaj$^1$}
\author{Sudhakar Yarlagadda$^{1,2}$}
\affiliation{${^1}$ CMP Division, Saha Institute of Nuclear Physics,
Kolkata, India}
\affiliation{${^2}$Cavendish Lab, Univ. of Cambridge, Cambridge, UK}
\date{\today}
\begin{abstract}
Studying the nature and consequences of electron-phonon interaction in manganites is an area of
intense ongoing research. 
Here, in an attempt to model charge and orbital ordering in
manganites displaying C-type antiferromaganetism,  we study cooperative Jahn-Teller effect in  two-band one-dimensional chains
in the regimes of both strong and weak electron-phonon couplings.
These chains exhibit orbital ferromagnetism with only $d_{z^2}$ orbitals being occupied.
At strong coupling and in the antiadiabatic regime, using a controlled analytic nonperturbative treatment that accounts for the quantum nature of the phonons,
we derive the effective polaronic Hamiltonians for a single chain as well as for interacting identically-long chains. Due to cooperative effects, 
these effective Hamiltonians manifest a
 dominant next-nearest-neighbor hopping compared to the usual nearest-neighbor hopping and a significantly enhanced nearest-neighbor repulsion.
 For densities up to half filling, 
upon tuning electron-phonon coupling, interacting-chain [single-chain] Jahn-Teller systems  
undergo quantum phase transition from a charge disordered state to a conducting charge-density-wave state characterized by a wavevector
$\vec{k} = (\pi ,\pi )$ [$k  = \pi $].
 On the other hand, up to half filling, a weak coupling analysis reveals a transition from a disordered state to an
insulating charge-density-wave state with a wavevector that depends linearly on the density;
the ordering is analyzed within a Peierls instability framework involving the dynamic noninteracting susceptibility
at nesting wavevector and phonon frequency.
Our analysis provides an opportunity to identify the regime of electron-phonon coupling in manganites through
experimentally determining the charge-ordering wavevector.
\end{abstract}
\maketitle
\section{Introduction}
Understanding the wide variety of exotic long-range orders in transition
metal oxides (such as manganites) 
and designing artificial structures  (such as heterostructures,  
quantum wires, and quantum dots) using these materials is of immense fundamental
interest  and also of huge technological importance.
Perovskite manganites R$_{1-x}$A$_x$MnO$_3$ (R = La, Pr, Nd, etc., A = Sr, Ca)
are  systems containing  ${\rm Mn}^{3+}$ ions with
one electron  in doubly degenerate $e_g$
orbitals; this results in cooperative Jahn-Teller (CJT) distortion (on adjacent sites)
 which lifts the orbital degeneracy and produces a cooperative
occupation of orbitals. Consequently, a strong interplay ensues between
charge, spin, and orbital degrees of freedom. 
This interplay 
leads to closely competing energy states with a variety of spin textures (such as
 metallic ferromagnets and
A-, C-, CE-, or G-type antiferromagnets), charge orders [such as density-dependent
and density-independent charge density waves (CDWs)], and orbital orders (such as ferromagnetic and antiferromagnetic
orbital density waves) \cite{cnr,khomskii,hotta}.

There have been numerous studies of the many-polaron effects
produced by quantum phonons in a one-band one-dimensional Holstein
model \cite{au:holstein} which is a simple case of noncooperative 
electron-phonon interaction \cite{au:sdadys,au:sdys,fradkin,capone,zheng,perroni,hamer,fehske}.
However, controlled mathematical
 modeling of CJT quantum
 systems 
that goes beyond modeling localized carriers  \cite{gehring}
has remained elusive (at least to our knowledge).
In fact, a controlled analytic treatment of  cooperative-breathing-mode effects 
in a single-band, many-polaron chain has  been  developed only recently \cite{au:rpys}.
The situation is also complicated 
 because of
lack of conclusive identification of the
strength of the electron-phonon interaction in oxides such as manganites.
Evidence of strong local Jahn-Teller distortions pointing to strong electron-phonon
coupling has been reported in manganites  by
 direct techniques such as EXAFS \cite{bianc},
 pulsed neutron diffraction 
\cite{louca},
 or through direct evidence of orbital ordering using
 resonant X-ray scattering 
\cite{murak}. On the other hand, weak coupling was inferred in the overdoped regime (i.e., $x > 0.5$)
using transmission electron microscopy \cite{mathur1,mathur3}, orientation-dependent transport measurements \cite{scox},
terahertz spectroscopy (to measure conductivity and permittivity) 
\cite{gorshunov,gorshunov2},
and  coherent synchrotron source (to measure optical conductivity) \cite{nucara}.

In the overdoped regime ($x>0.5$) of manganites such as La$_{1-x}$Ca$_x$MnO$_3$, at low temperatures,
charge ordering results along with orbital ordering and antiferromagnetism.
The nature of the charge order has been a subject of much debate.
Earlier on, charge order was described in terms of a strong-coupling picture of
stripes of localized charges where
the mixed-valence  manganese ions split into ${\rm Mn}^{3+}$
ions and ${\rm Mn}^{4+}$ ions \cite{goodenough2,chen}. However, based on the findings in Refs. \onlinecite{chen2,au:larochelle,mathur2},
a phenomenological picture of an extended CDW (instead of localized charges) was developed in Ref. \onlinecite{au:peter_nature}.
The experiments indicating weak coupling 
\cite{mathur1,mathur3,scox,nucara}, for $0.5 \leq x \leq 0.75$ in La$_{1-x}$Ca$_x$MnO$_3$,
also seem to agree with this picture of weakly-modulated CDW characterized by a concentration-dependent wavevector $q=(1-x)a^*$ with $a^*$ being
the reciprocal lattice vector \cite{gorsh_quest}. 
Furthermore, in another manganite (i.e., La$_{0.2}$Sr$_{0.8}$MnO$_3$),
 charge ordering wavevector $q=0.2 a^*$  has  been reported \cite{maiti2}.
As regards magnetic ordering, in the overdoped regimes of
La$_{1-x}$Sr$_x$MnO$_3$ (for $0.65 < x<0.95$) \cite{au:chmaissem,au:szewczyk,maiti1}   
and  La$_{1-x}$Ca$_x$MnO$_3$ (for $0.75 \leq x \leq 0.87$), \cite{yao,au:kallias, au:goodenough} C-type antiferromaganetism
has also been reported.

In this paper, compared to the phenomenological theory of Ref. \onlinecite{au:peter_nature},
we attempt at a microscopic theory in the overdoped regime.
To reduce the daunting complexity, we assume C-type antiferromagnetism
and invoke only the charge and orbital degrees of freedom
to model the remaining possibilities of charge and orbital orderings.
We investigate the regimes of weak coupling and strong coupling and show how the regime of coupling can be ascertained
experimentally.
Since manganites with C-type antiferromaganetic 
ordering can be effectively considered as chains,
we study the
effect of electron-phonon coupling in a single chain and in
interacting chains.
At strong coupling, we find that the charge ordering wavevector is
always $q=0.5a^*$ irrespective of the value of the doping $x$; whereas at weak coupling, the ordering wavevector
has the density-dependent form $q=(1-x)a^*$. Thus, based on the experimental determination
of the charge ordering wavevectors at various doping values $x$ in C-type antiferromagnets,
we believe that our theory offers an 
opportunity for 
identifying
 the regime of electron-phonon coupling.

The paper is organized as follows. We derive a general Hamiltonian for CJT effect
in three dimensions in Sec.~II. In Sec.~III, we derive the effective Hamiltonian for
CJT interaction in one dimension and study its nature 
at various electron-phonon couplings. In the subsequent  Secs. IV  and V,
we analyze the CJT effect in interacting chains 
for the case of strong coupling 
in the antiadiabatic regime and for the case of  weak coupling in the adiabatic regime, respectively.
Before closing, we discuss the connection between our system and other systems in Sec.~VI and present
our conclusions in Sec.~VII. 
\section{General CJT Hamiltonian}
The general Hamiltonian for the CJT system in manganites 
can be written as 
$H^G=H_t+H_{ep}+H_l$,
where $H_t$ is the hopping term, $H_{ep}$ the electron-phonon-interaction term,
and $H_l$ the lattice term.
We start with an over-complete basis
$\psi_x = 3x^2-r^2, \psi_y = 3y^2-r^2, \psi_z = 3z^2-r^2$ which satisfies the relation
$\psi_x+\psi_y+\psi_z=0$.
 The basis state $\psi_z$ corresponds to the $d_{z^2}$ orbital
depicted in Fig.\ref{fig:cartoon}.
The hopping term can be expressed in the above basis as:
\begin{IEEEeqnarray}{rCl}
H_t  = &-&t\sum_{i,j,k}[\{d^\dagger_{x^2;i+1,j,k}d_{x^2;i,j,k}+d^\dagger_{y^2;i,j+1,k}d_{y^2;i,j,k} \nonumber \\
& + & d^\dagger_{z^2;i,j,k+1}d_{z^2;i,j,k}\} + {\rm H.c.}], 
\label{eq:hopover}
\end{IEEEeqnarray}
where $ d^\dagger_{x^2;i,j,k},d^\dagger_{y^2;i,j,k}, d^\dagger_{z^2;i,j,k} $ 
 are creation 
 operators
at the site $(i,j,k)$  for $d_{x^2}$, $d_{y^2}$, and $d_{z^2}$ orbitals, respectively. The labelling indices $i$, $j$, and $k$
run along the $x$-, $y$-, and $z$-axes, respectively.
The electron-phonon interaction term can be written as:
\begin{IEEEeqnarray}{rCl}
H_{ep}  = & -& g \omega_0 \sqrt{2 M \omega_0} \sum_{i,j,k} [n_{x^2;i,j,k} Q_{x;i,j,k} \nonumber \\
& + & n_{y^2;i,j,k} Q_{y;i,j,k} + n_{z^2;i,j,k} Q_{z;i,j,k} ],
\label{eq:elphover}
\end{IEEEeqnarray}
where $g$ is the electron-phonon coupling, $M$ is the mass of an oxygen ion,  $\omega_0$ is the frequency of optical phonons, 
and $n_{x^2(y^2,z^2);i,j,k} = d^\dagger_{x^2(y^2,z^2);i,j,k} d_{x^2(y^2,z^2);i,j,k} $ are the number operators.
Furthermore, $Q_{x;i,j,k}$, $Q_{y;i,j,k}$ and $Q_{z;i,j,k}$ are defined in terms of the displacements
[$u_{x;i,j,k}~ \& ~u_{x;i-1,j,k}$; $ u_{y;i,j,k}~ \&~ u_{y;i,j-1,k}$; $ u_{z;i,j,k} ~\&~ u_{z;i,j,k-1}$]
of oxygen ions 
around (and in the direction of) the $d_{x^2}$, $d_{y^2}$, and $d_{z^2} $ orbitals, respectively, as follows:
$Q_{x;i,j,k} = u_{x;i,j,k}-u_{x;i-1,j,k}$, $Q_{y;i,j,k} = u_{y;i,j,k}-u_{y;i,j-1,k}$, and $Q_{z;i,j,k} = u_{z;i,j,k}-u_{z;i,j,k-1}$.
Here, besides considering the displacement of the ions, we also consider their kinetic energy, thereby 
invoking quantum nature of the phonons. Then,
the lattice Hamiltonian is given by
\begin{IEEEeqnarray}{rCl}
H_l  = && \frac{M}{2} \sum_{i,j,k} [ \dot{u}^2_{x;i,j,k} + \dot{u}^2_{y;i,j,k} + \dot{u}^2_{z;i,j,k} ] \nonumber \\ 
& + & \frac{K}{2} \sum_{i,j,k} [ u^2_{x;i,j,k} + u^2_{y;i,j,k} + u^2_{z;i,j,k} ], 
\label{eq:gen_lattice}
\end{IEEEeqnarray}
where $ \dot{u}_{x;i,j,k} $, $ \dot{u}_{y;i,j,k} $, and $ \dot{u}_{z;i,j,k} $ are the time derivatives
 of the oxygen-ion displacements 
$u_{x;i,j,k}$, $ u_{y;i,j,k}$, and $ u_{z;i,j,k}$, respectively.

 The usual orthogonal basis states $\psi_{x^2-y^2}$ and $\psi_{z^2}$ are related to the over-complete basis states
$\psi_x$, $\psi_y$, and $\psi_z$ as follows:
\begin{IEEEeqnarray*}{rCl}
&&\psi_{x^2-y^2} = \frac{1}{\sqrt3} (\psi_x-\psi_y), \\
&&\psi_{z^2} = \psi_z.
\label{eq:psi_x}
\\*\IEEEyesnumber
\end{IEEEeqnarray*}
From Eq.~\eqref{eq:psi_x} we get,
\begin{IEEEeqnarray}{rCl}
\psi_x &=& \frac{1}{2} (\sqrt{3} \psi_{x^2-y^2}-\psi_{z^2}), \nonumber \\
\psi_y &=& - \frac{1}{2} (\sqrt{3} \psi_{x^2-y^2}+\psi_{z^2}), \nonumber \\
\psi_z &=& \psi_{z^2}. \nonumber\\
\label{eq:psiz^2}
\end{IEEEeqnarray}

 Next, using Eq.~\eqref{eq:psiz^2}, we express the general Hamiltonian in  the  orthogonal basis $\psi_{x^2-y^2}$ and $\psi_{z^2}$ as follows:
\begin{widetext}
\begin{IEEEeqnarray}{rCl}
\label{eq:gen_hop_com}
H_t = 
&-&\frac{t}{4} \sum_{i,j,k} \{(d^\dagger_{z^2;i+1,j,k},d^\dagger_{x^2-y^2;i+1,j,k}) 
\begin{pmatrix}
1 & -\sqrt{3} \\
-\sqrt{3} & 3
\end{pmatrix}
\begin{pmatrix}
d_{z^2;i,j,k} \\
d_{x^2-y^2;i,j,k}        
\end{pmatrix} 
+ {\rm H. c.} \}  
-\frac{t}{4} \sum_{i,j,k} \{ (d^\dagger_{z^2;i,j+1,k},d^\dagger_{x^2-y^2;i,j+1,k}) \nonumber \\
&\times&
\begin{pmatrix}
1 & \sqrt{3} \\
\sqrt{3} & 3
\end{pmatrix}
\begin{pmatrix}
d_{z^2;i,j,k} \\
d_{x^2-y^2;i,j,k}        
\end{pmatrix} 
+ {\rm H. c.} \} 
-t\sum_{i,j,k} \{(d^\dagger_{z^2;i,j,k+1},d^\dagger_{x^2-y^2;i,j,k+1}) 
\begin{pmatrix}
1 & 0 \\
0 & 0
\end{pmatrix} 
\begin{pmatrix}
d_{z^2;i,j,k} \\
d_{x^2-y^2;i,j,k}        
\end{pmatrix}
+ {\rm H. c.} \},
\end{IEEEeqnarray}
\begin{IEEEeqnarray}{rCl}
H_{ep} = &-&\frac{1}{4} g \omega_0 \sqrt{2 M \omega_0} \nonumber \\
&\times&\sum_{i,j,k} (d^\dagger_{z^2;i,j,k},d^\dagger_{x^2-y^2;i,j,k}) 
\begin{pmatrix}
Q_{x;i,j,k}+Q_{y;i,j,k}+4Q_{z;i,j,k} & -\sqrt{3}Q_{x;i,j,k}+\sqrt{3}Q_{y;i,j,k} \\
-\sqrt{3}Q_{x;i,j,k}+\sqrt{3}Q_{y;i,j,k} & 3Q_{x;i,j,k}+3Q_{y;i,j,k}
\end{pmatrix}
\begin{pmatrix}
d_{z^2;i,j,k} \\
d_{x^2-y^2;i,j,k}        
\end{pmatrix} ,
\label{eq:gen_elph_com}
\end{IEEEeqnarray}
and $H_l$ is again given by Eq. (\ref{eq:gen_lattice}).
Here, it should be mentioned that an expression for $H^G$ in an alternate basis has been derived  in Ref. \onlinecite{au:allen}; 
however, these authors consider classical phonons. 

In the subsequent sections, we shall study special cases of our above general CJT Hamiltonian, 
namely, the single chain and the interacting chains.

\begin{figure}
\begin{center}
\vspace{-1cm}
\includegraphics[]{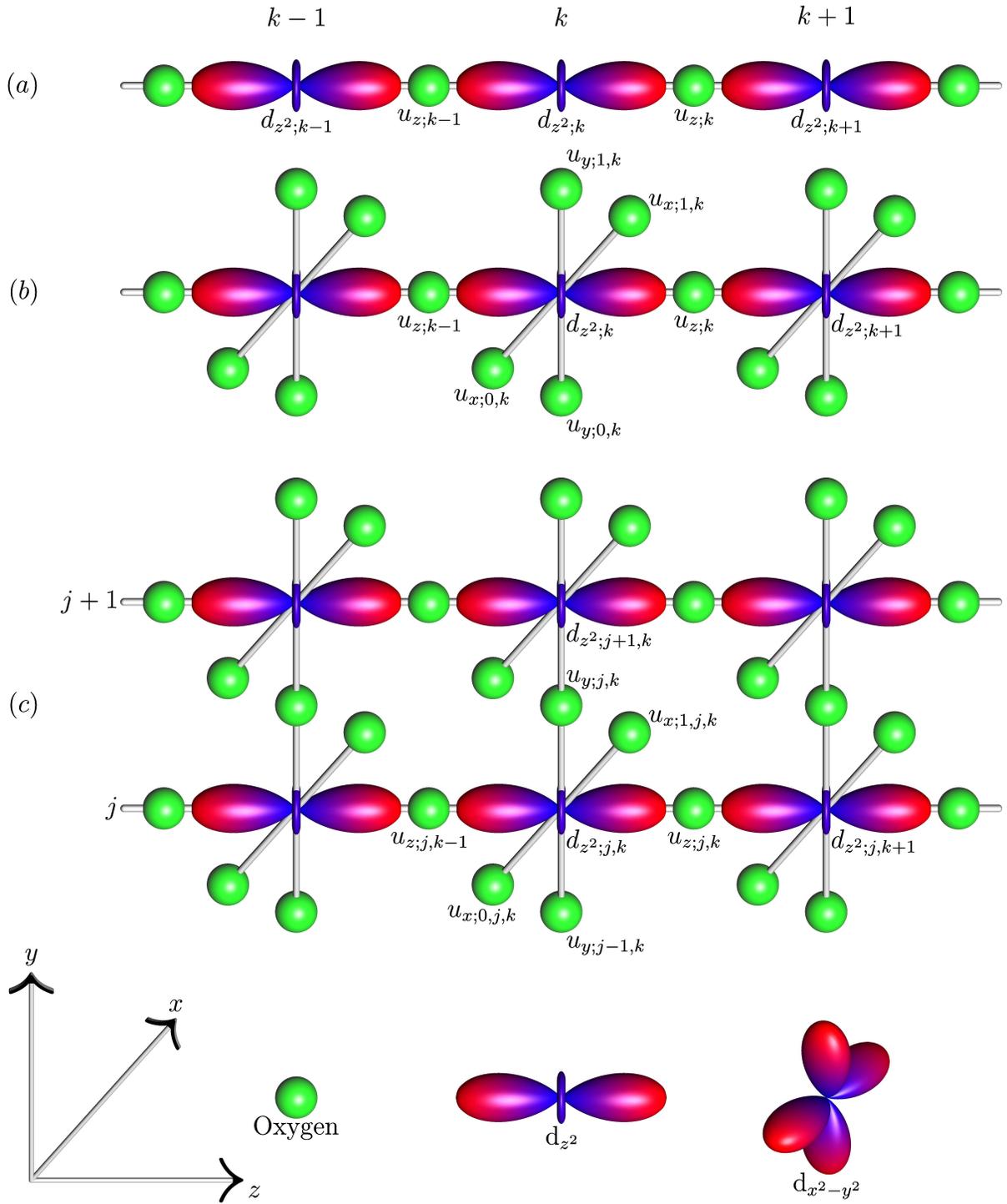}
\caption{(Color online) Depiction of $(a)$ one-dimensional single-band chain with cooperative breathing mode; $(b)$  cooperative  Jahn-Teller chain
involving $d_{z^2}$ and $d_{x^2-y^2}$ orbitals;
and $(c)$ interacting  cooperative Jahn-Teller chains with $d_{z^2}$ and $d_{x^2-y^2}$ orbitals. For simplicity, only  $d_{z^2}$ orbitals
are displayed in (b) and (c).}
\label{fig:cartoon}
\end{center}
\end{figure}
\end{widetext}

\section{Single chain CJT model at strong coupling}

We consider a one-dimensional  Jahn-Teller chain with cooperative electron-phonon 
interaction along the z-direction and non-cooperative  electron-phonon interaction (of the Holstein-type \cite{au:holstein,au:sdadys})
along the x- and
y-directions as shown in Fig.~\ref{fig:cartoon}$(b)$. The lattice term given by Eq.~\eqref{eq:gen_lattice} can be 
written for this case as follows:
\begin{IEEEeqnarray*}{rCl}
H^{CJT}_l  = &&\> \frac{M}{2} \sum_{k} [\dot{u}^2_{x;0,k}+\dot{u}^2_{x;1,k} + \dot{u}^2_{y;0,k}+\dot{u}^2_{y;1,k}\\
&&\>+ \dot{u}^2_{z;k}]
+\frac{K}{2} \sum_{k} [ u^2_{x;0,k}+u^2_{x;1,k} + u^2_{y;0,k}\\
&&\>+u^2_{y;1,k} + u^2_{z;k} ].\IEEEyesnumber 
\label{eq:1dlatticeintermediate}
\end{IEEEeqnarray*}
We define $Q^{\prime}_{x;k} \equiv u_{x;1,k}+u_{x;0,k}$, $Q^{\prime}_{y;k} \equiv u_{y;1,k}+u_{y;0,k}$,
$Q_{x;k} \equiv u_{x;1,k}-u_{x;0,k}$, and $Q_{y;k} \equiv u_{y;1,k}-u_{y;0,k}$ 
and incorporate these definitions in 
Eq.~\eqref{eq:1dlatticeintermediate} to obtain
\begin{IEEEeqnarray*}{rCl}
&& \!\! H^{CJT}_l \\
&& ~ =  \frac{M}{2} \sum_{k} \bigg [ \frac{1}{2}\{\dot{Q}^{\prime 2}_{x;k}+\dot{Q}^{\prime 2}_{y;k}\}
+\frac{1}{4}\{\dot{Q}^{+2}_{xy;k} 
 + \dot{Q}^{-2}_{xy;k}\}+ \dot{u}^2_{z;k} \bigg ] \\
&& ~~ +\frac{K}{2} \sum_{k} \bigg [\frac{1}{2} \{Q^{\prime 2}_{x;k}+Q^{\prime 2}_{y;k}\} 
 + \frac{1}{4}\{Q^{+2}_{xy;k}+Q^{-2}_{xy;k}\}+u^2_{z;k} \bigg ], \\
\IEEEyesnumber 
\label{eq:1dlattice}
\end{IEEEeqnarray*}
where $Q^\pm_{xy;k} \equiv Q_{x;k} \pm Q_{y;k} $.
For the  present single chain case,  Eqs.~\eqref{eq:gen_hop_com} and
\eqref{eq:gen_elph_com} reduce to the following equations:
\begin{IEEEeqnarray*}{rCl}
H^{CJT}_t = &-&t \sum_k (d^\dagger_{z^2;k+1}d_{z^2;k}+ {\rm H.c.}) ,
\IEEEyesnumber
\label{eq:HCJT_t}
\end{IEEEeqnarray*}
and 
\begin{IEEEeqnarray*}{rCl}
&&\!\!\!\! \frac{H^{CJT}_{ep}}{g \omega_0 \sqrt{2 M \omega_0} } \\
&&  = -\sum_k \bigg [ \Big \{(u_{z;k}-u_{z;k-1})
+\frac{1}{4}Q^+_{xy;k} \Big \} d^\dagger_{z^2;k}d_{z^2;k} \\
&& \qquad \qquad + \frac{3}{4}Q^+_{xy;k} d^\dagger_{x^2-y^2;k}d_{x^2-y^2;k} \\
&& \qquad \qquad -\frac{\sqrt{3}}{4} Q^-_{xy;k} \Big ( d^\dagger_{z^2;k}d_{x^2-y^2;k} + {\rm H.c.} \Big ) \bigg ] . 
\IEEEyesnumber
\label{eq:HCJT_ep}
\end{IEEEeqnarray*}
Next, we note that the center-of-mass displacement terms $Q^{\prime }_{x;k}$ and $Q^{\prime }_{y;k}$ as well as
the center-of-mass momentum terms $\dot{Q}^{\prime }_{x;k}$ and $\dot{Q}^{\prime }_{y;k}$ 
 of Eq. \eqref{eq:1dlatticeintermediate}
do not couple to the electrons 
 [as can be seen from Eqs. (\ref{eq:HCJT_t}) and \eqref{eq:HCJT_ep}]. Hence, for our single chain case, 
Eq. (\ref{eq:1dlattice}) simplifies to be
\begin{IEEEeqnarray*}{rCl}
H^{CJT}_l &=& 
\sum_k\left[\frac{1}{2} M  {{\dot{u}}_{z;k}}^2 + \frac{1}{2} K  u^2_{z;k}\right] \\
&& +\sum_k\left[\frac{1}{2}\frac{M}{4} {{\dot{Q}^{+2}}_{xy;k}} + \frac{1}{2}\frac{K}{4} Q^{+2}_{xy;k}\right]  \\
&& + \sum_k\left[\frac{1}{2}\frac{M}{4}  {\dot{Q}^{-2}_{xy;k}} + \frac{1}{2}\frac{K}{4} Q^{-2}_{xy;k}\right] .
\IEEEyesnumber\label{eq:HCJT_l}
\end{IEEEeqnarray*}
The general Hamiltonian for the present single chain CJT case can be expressed as follows by adding Eqs. \eqref{eq:HCJT_t}, \eqref{eq:HCJT_ep}, and
\eqref{eq:HCJT_l}:
\begin{IEEEeqnarray}{rCl}
 H^{CJT}=H^{CJT}_t+H^{CJT}_{ep}+H^{CJT}_{l} .
\label{eq:HCJT}
\end{IEEEeqnarray}
 Next, by using the following second-quantized representation of the various displacement operators:
\begin{IEEEeqnarray*}{rCl}
u_{z;k} = \frac{a^\dagger_{z;k}+a_{z;k}}{\sqrt{2M\omega_0}},
Q^+_{xy;k}= \frac{b^\dagger_k+b_k}{\sqrt{2\frac{M}{4}\omega_0}},
Q^-_{xy;k}= \frac{c^\dagger_k+c_k}{\sqrt{2\frac{M}{4}\omega_0}},
\end{IEEEeqnarray*}
in the above Hamiltonian of Eq. (\ref{eq:HCJT}), we obtain
\begin{IEEEeqnarray}{rCl}
\!\!\!\!\!\!\!\! H^{CJT} =& -&t \sum_k (d^\dagger_{z^2;k+1}d_{z^2;k}+ {\rm H.c.}) \nonumber \\ 
&-&g \omega_0 \sum_k \bigg [(a^\dagger_{z;k}+a_{z;k})(n_{z^2;k}-n_{z^2;k+1}) \nonumber \\
&&\qquad ~~~~+  \frac{1}{2} (b^\dagger_k+b_k) (n_{z^2;k}+3 n_{x^2-y^2;k} ) \nonumber \\
&& \qquad ~~~~ - \frac{\sqrt{3}}{2} (c^\dagger_k+c_k) (d^\dagger_{z^2;k}d_{x^2-y^2;k}+ {\rm H.c.})\bigg ] \nonumber \\
&+& \omega_0 \sum_k (a^\dagger_{z;k}a_{z;k} + b^\dagger_kb_k + c^\dagger_kc_k), 
\label{eq:HCJT_sq}
\end{IEEEeqnarray}
where $n_{z^2;k} \equiv d^\dagger_{z^2;k}d_{z^2;k} $ and $n_{x^2-y^2;k} \equiv d^\dagger_{x^2-y^2;k}d_{x^2-y^2;k} $.

We will now adapt the well-known Lang-Firsov transformation \cite{lang} for the above Hamiltonian so that we can
perform perturbation in the polaronic (Lang-Firsov) frame of reference. The transformed Hamiltonian is given by $\tilde{H}^{CJT} = \exp(S) H^{CJT} \exp(-S)$
where
\begin{IEEEeqnarray}{rCl}
S =& -&g \sum_k [(a^\dagger_{z;k}-a_{z;k})(n_{z^2;k}-n_{z^2;k+1}) \nonumber \\
&+& \frac{1}{2} (b^\dagger_k-b_k) (n_{z^2;k}+3 n_{x^2-y^2;k} )].  
\end{IEEEeqnarray}
Here, in our modified Lang-Firsov transformation, it should be noted that we have included only the density terms 
and ignored the orbital-flip terms  ($d^\dagger_{z^2;k}d_{x^2-y^2;k}$ and its Hermitian conjugate) appearing in the interaction
part of the above equation (\ref{eq:HCJT_sq}). This choice is dictated by mathematical expediency
to arrive at an analytic expression. 
Then, the Lang-Firsov transformed Hamiltonian is given by $\tilde{H}^{CJT} = H_0 + H_1$ where
\begin{IEEEeqnarray}{rCl}
H_0 &=& \omega_0 \sum_k (a^\dagger_{z;k}a_{z;k} + b^\dagger_kb_k + c^\dagger_kc_k) \nonumber \\
&&- \frac{9}{4} g^2 \omega_0 \sum_k (n_{z^2;k}  
+ n_{x^2-y^2;k}) \nonumber  \\
&&-\frac{3}{2} g^2 \omega_0 \sum_k n_{z^2;k} n_{x^2-y^2;k} 
+ 2 g^2 \omega_0 \sum_k n_{z^2;k}n_{z^2;k+1} \nonumber \\   
&&- t e^{-\frac{13}{4}g^2} \sum_k (d^\dagger_{z^2;k+1}d_{z^2;k}+ {\rm H.c.}),
\label{eq:H_0}
\end{IEEEeqnarray}
where the term $2 g^2 \omega_0 \sum_k n_{z^2;k}n_{z^2;k+1}$ arises because of the cooperative nature of the interaction; furthermore,
the attractive interaction term $-\frac{3}{2} g^2 \omega_0 \sum_k n_{z^2;k} n_{x^2-y^2;k}$ will be negated
by a much larger repulsive Coulombic term  $U \sum_k n_{z^2;k} n_{x^2-y^2;k}$ because of which no site can have both the orbitals 
occupied simultaneously. The remaining term of $\tilde{H}^{CJT}$ is given by
 $H_1 \equiv H^I_1+ H^{II}_1$ with 
\begin{IEEEeqnarray}{rCl}
H^I_1 
= -t e^{-\frac{13}{4} g^2}
 \sum_k [ d^\dagger_{z^2;k+1}d_{z^2;k}
\{
{\cal{T}}_{+}^{k \dagger} {\cal{T}}^{k}_{-}
-1 \} + {\rm H.c.}],
\IEEEeqnarraynumspace
\label{eq:HI}
\end{IEEEeqnarray}
where ${\cal{T}}^{k}_{\pm} \equiv \exp[\pm g( 2 a_{z;k} - a_{z;k-1} - a_{z;k+1})\pm \frac{g}{2}(b_k-b_{k+1})]$ 
and 
\begin{IEEEeqnarray}{rCl}
H^{II}_1 
&=& \frac{\sqrt{3}}{2} g \omega_0 e^{-\frac{3}{2}g^2} \sum_k (c^\dagger_k+c_k)\bigg [d^\dagger_{z^2;k}d_{x^2-y^2;k}
 \nonumber \\
&&\times\>e^{g(a^\dagger_{z;k-1}-a^\dagger_{z;k}+b^\dagger_k)} e^{-g(a_{z;k-1}-a_{z;k}+b_k)} + {\rm H.c.}\bigg].
\IEEEeqnarraynumspace
\label{eq:HII}
\end{IEEEeqnarray}
Now, to perform perturbation theory, we note that the eigenstates of $H_0$ are given by $|n,m\rangle =|n\rangle_{el}\otimes |m\rangle_{ph}$
with $|0,0\rangle$ being the ground state. We consider the strong-coupling case $g^2 >> 1$ and the antiadiabatic regime $t/\omega_0 <1$; consequently,
the coefficients of the perturbation terms $H^I$ and $H^{II}$ in Eqs. (\ref{eq:HI}) and (\ref{eq:HII}), respectively, 
satisfy the conditions $t e^{-\frac{13}{4} g^2} << \omega_0$ 
and $\frac{\sqrt{3}}{2} g \omega_0 e^{-\frac{3}{2}g^2} << \omega_0$. Now, the 
second-order perturbation term [obtained using Schrieffer-Wolff transformation
 as mentioned in Eq. (6) of Ref. \onlinecite{au:rpys}]
is expressed as
\begin{IEEEeqnarray}{rCl}
H^{(2)} = \sum_{m}
\frac{\langle 0|_{ph} H_{1} |m\rangle_{ph}
 \langle m|_{ph} H_{1} |0\rangle_{ph}}
{E_{0}^{ph} - E_{m}^{ph}} .
\label{H^2}
 \end{IEEEeqnarray}   
In  Eq.~\eqref{H^2}, the contribution of cross terms involving  $H^I_1$ and $H^{II}_1$ is zero because the phonons 
do not match; hence, we get 
\begin{IEEEeqnarray}{rCl}
H^{(2)} =&& \sum_{m}
\frac{\langle 0|_{ph} H^I_{1} |m\rangle_{ph}
 \langle m|_{ph} H^I_{1} |0\rangle_{ph}}
{E_{0}^{ph} - E_{m}^{ph}} \nonumber \\
&+&\sum_{m} \frac{\langle 0|_{ph} H^{II}_{1} |m\rangle_{ph}
 \langle m|_{ph} H^{II}_{1} |0\rangle_{ph}}
{E_{0}^{ph} - E_{m}^{ph}}.
\label{H^2_1}
 \end{IEEEeqnarray} 
We will first evaluate the term involving $H^{II}_1$ in the above equation. After some algebra, we get the following expression:
 \begin{IEEEeqnarray}{rCl}
&& \!\!\!\! \sum_{m} \frac{\langle 0|_{ph} H^{II}_{1} |m\rangle_{ph}
 \langle m|_{ph} H^{II}_{1} |0\rangle_{ph}}
{E_{0}^{ph} - E_{m}^{ph}} \nonumber \\
&& \approx -\frac{\omega_0}{4} \sum_k \Big [ n_{z^2;k}  
+ n_{x^2-y^2;k} 
- 2 n_{z^2;k} n_{x^2-y^2;k} \Big ] .
\label{H^II}
 \end{IEEEeqnarray} 
We note that the coefficients of the  terms $n_{z^2;k}$, $ n_{x^2-y^2;k}$, and $ n_{z^2;k} n_{x^2-y^2;k}$
in the above equation are much smaller than the coefficients of the same terms in Eq. (\ref{eq:H_0}); consequently, we ignore the contribution from 
Eq. (\ref{H^II}) in the expression for the effective Hamiltonian of the CJT chain.  
 
Next, after performing some tedious algebra (using considerations similar to those in Ref. \onlinecite{au:rpys}), the effective Hamiltonian can be obtained as:
\begin{IEEEeqnarray*}{rCl}
 H^{CJT}_{eff} 
&=&-t e^{-\frac{13}{4} g^2}
 \sum_k (d^\dagger_{z^2;k+1}d_{z^2;k}+ {\rm H.c.} ) \\ 
&&~-\frac{t^2}{\omega_0}e^{-\frac{13}{2} g^2}
G_3 \left(2,2,\frac{1}{4}\right) \\
&& \qquad \times \sum_k \left [d^\dagger_{z^2;k-1}(1-2n_{z^2;k})d_{z^2;k+1}
+ {\rm H.c.}\right ] \\
&&~+2\Bigg[g^2\omega_0+\frac{t^2}{\omega_0}e^{-\frac{13}{2} g^2}
 G_5 \left(4,1,1,\frac{1}{4},\frac{1}{4}\right)
 \Bigg] \\
&& \qquad \times 
\sum_k n_{z^2;k} n_{z^2;k+1} ,
\IEEEyesnumber
\label{eq:1djtseries}
\end{IEEEeqnarray*}
where  
\begin{IEEEeqnarray*}{rCl}
G_3 \left(2,2,\frac{1}{4}\right) &\equiv& F_3 \left(2,2,\frac{1}{4}\right)
+F_2 \left(2,2\right) 
+2F_2 \left(2,\frac{1}{4}\right) \\
&&~+2F_1 \left(2\right)+F_1 \left(\frac{1}{4}\right) ,
\IEEEyesnumber
\end{IEEEeqnarray*}
and
\begin{IEEEeqnarray*}{rCl}
&& G_5 \left(4,1,1,\frac{1}{4},\frac{1}{4}\right) \\ 
&&\equiv
 F_5 \left(4,1,1,\frac{1}{4},\frac{1}{4}\right)
+2F_4 \left(4,1,1,\frac{1}{4}\right) 
 +2F_4 \left(4,1,\frac{1}{4},\frac{1}{4}\right) \\
&& ~+F_4 \left(1,1,\frac{1}{4},\frac{1}{4}\right)
+F_3 \left(4,1,1\right)+4F_3 \left(4,1,\frac{1}{4}\right) \\
&& ~+F_3 \left(4,\frac{1}{4},\frac{1}{4}\right)
+2F_3 \left(1,1,\frac{1}{4}\right)
+2F_3 \left(1,\frac{1}{4},\frac{1}{4}\right) \\
&& ~ +2F_2 \left(4,1\right)
+2F_2 \left(4,\frac{1}{4}\right) 
+F_2 \left(1,1\right) 
 +4F_2 \left(1,\frac{1}{4}\right) \\
&& ~ +F_2 \left(\frac{1}{4},\frac{1}{4}\right)
+F_1 \left(4\right) 
 +2F_1 \left(1\right)
+2F_1 \left(\frac{1}{4}\right),
\IEEEyesnumber
\end{IEEEeqnarray*}
with 
\begin{IEEEeqnarray*}{rCl}
F_n(\alpha_1, \ldots , \alpha_n ) \equiv \sum_{m_1=1}^{\infty} 
 ...
\sum_{m_n=1}^{\infty}
 \frac {(\alpha_1 g^2)^{m_1} ... (\alpha_n g^2)^{m_n}}
{m_1!\ldots m_n!(m_1+ \ldots + m_n)}.
 \end{IEEEeqnarray*}   
A general term of the form $G_n(\alpha_1,\alpha_2,\ldots,\alpha_n)$ 
can be expressed as $G_n(\alpha_1,\alpha_2,\ldots,\alpha_n) = F_n(\alpha_1,\alpha_2,\ldots,\alpha_n)+\sum_{k=1}^{n-1}\sum_c
F_k(\alpha_{c_1},\alpha_{c_2},\ldots,\alpha_{c_k})$ where the summation over $c$ represents summing over all possible
$^nC_m$ combinations of $m$ arguments chosen from the total set of 
 $n$ arguments $\{\alpha_1,\alpha_2,\ldots,\alpha_n\}$. We then obtain the following useful relationship
(derived in Appendix \ref{app:serieses}):
\begin{IEEEeqnarray*}{rCl}
&& \!\!\!\!\!\! G_n(\alpha_1,\alpha_2,\ldots,\alpha_n) \\
&&= F_n(\alpha_1,\alpha_2,\ldots,\alpha_n)+\sum_{k=1}^{n-1}\sum_c
F_k(\alpha_{c_1},\alpha_{c_2},\ldots,\alpha_{c_k})\\
\qquad&&=\int \frac{e^{\sum_{i=1}^n\alpha_ig^2}-1}{g^2}dg^2
= \sum^\infty_{m=1} \frac{(\sum_{i=1}^n\alpha_ig^2)^m}{m\>m!}\\
\qquad&&
\stackrel{\text{for}\>g^2 >>1}{\approx}
\frac{e^{\sum_{i=1}^n\alpha_ig^2}}{\sum_{i=1}^n\alpha_ig^2}  .
\IEEEyesnumber
\label{eq:Gsimpl}
\end{IEEEeqnarray*}
Then, on using the above approximation for $G_n(\alpha_1,\alpha_2,\ldots,\alpha_n)$
at large $g^2$,  Eq. (\ref{eq:1djtseries}) simplifies as follows: 
\begin{IEEEeqnarray*}{rCl}
&&\!\!\!\!\!\!\!\!\!\! \frac{H^{CJT}_{eff}}{te^{-\frac{13}{4}g^2}} \\
& \approx & -\sum_k (d^\dagger_{z^2;k+1}d_{z^2;k}+ {\rm H.c.} )\\
&&-\frac{4}{17}\frac{te^{g^2}}{g^2\omega_0}\sum_k \left[d^\dagger_{z^2;k-1}(1-2n_{z^2;k})d_{z^2;k+1}
+ {\rm H.c.}\right] \\
&&+ \left[ \frac{2g^2\omega_0}{t}  
+ \frac{4}{13} \frac{t}{g^2\omega_0}\right]e^{\frac{13}{4}g^2} \sum_k n_{z^2;k} n_{z^2;k+1} .
\IEEEyesnumber
\label{eq:1djteff}
\end{IEEEeqnarray*}
  
In contrast to the  effective Hamiltonian of Eq. (\ref{eq:1djtseries}), the relatively simpler case of 
single-band cooperative
breathing mode  [shown in Fig.~\ref{fig:cartoon}$(a)$] yields the following effective Hamiltonian\cite{au:rpys}:
\begin{IEEEeqnarray*}{rCl}
&&\!\!\!\!\!\!\!\! H^{CBM}_{eff} \\
=&& -t e^{-3g^2} \sum_k (d^\dagger_{z^2;k+1}d_{z^2;k}+ {\rm H.c.} ) \\  
&&-\frac{t^2}{\omega_0}e^{-6g^2}
G_2(2,2)
\sum_k \left [d^\dagger_{z^2;k-1}(1-2n_{z^2;k})d_{z^2;k+1}
+ {\rm H.c.}\right ]\\
&&+ 2\bigg [g^2\omega_0+\frac{t^2}{\omega_0}e^{-6g^2}
G_3(4,1,1)
\bigg]
 \sum_k n_{z^2;k} n_{z^2;k+1}.
\IEEEyesnumber
\label{eq:1dcbmseries}
\end{IEEEeqnarray*}
On using the relationship of Eq. (\ref{eq:Gsimpl}), the effective Hamiltonian for  cooperative-breathing-mode chain
[given by  Eq.~\eqref{eq:1dcbmseries}] simplifies to be
\begin{IEEEeqnarray*}{rCl}
&&\!\!\!\!\!\!\!\!\!\! \frac{H^{CBM}_{eff}}{te^{-3g^2}}\\
&\approx&-\sum_k(d^\dagger_{z^2;k+1}d_{z^2;k}+ {\rm H.c.})\\
&&-\frac{te^{g^2}}{4g^2\omega_0}\sum_k\left[d^\dagger_{z^2;k-1}(1-2n_{z^2;k})d_{z^2;k+1}+\text{H.c.}\right]\\
&&+\left[\frac{2g^2\omega_0}{t}+\frac{t}{3g^2\omega_0}\right]e^{3g^2}\sum_kn_{z^2;k} n_{z^2;k+1} .
\IEEEyesnumber
\label{eq:1dcbmeff}
\end{IEEEeqnarray*}

\begin{figure}
\includegraphics[height=7.5cm,width=6.5cm,angle=-90]{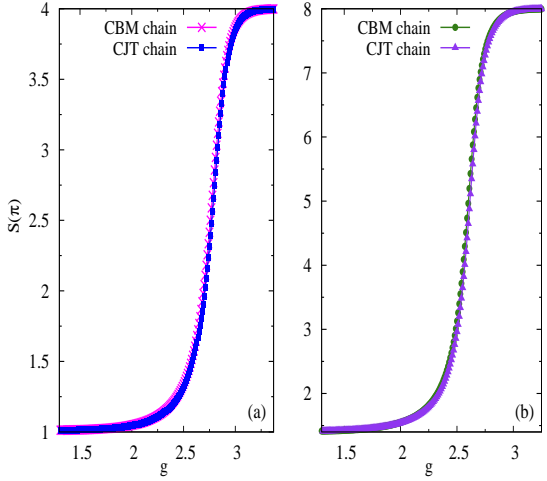} 
\caption{(Color online) Structure factor $S(\pi)$ showing CDW transition in CBM chain and CJT chain when
 $\frac{t}{\omega_0}=0.1$ and for $(a) N=16,~ N_p = 4$; and $(b) N=18,~ N_p = 6$. }
\label{fig:cbmjtcomp}
\end{figure}

Comparing Eq.~\eqref{eq:1djteff} with Eq.~\eqref{eq:1dcbmeff}, we find that the coefficients of next-nearest-neighbor (NNN)
hopping terms are approximately equal and so are the coefficients of the nearest-neighbor (NN) repulsion terms.
 Hence, both the CBM chain and the CJT chain should exhibit
similar behavior up to half-filling. We diagonalize these effective Hamiltonians using a modified Lanczos algorithm \cite{au:gagliano}
[with antiperiodic (periodic) boundary
conditions for even (odd) number of fermions]
and calculate the structure factor $S(k)$ at the ordering wavevector $\pi$.
Upon tuning the electron-phonon coupling $g$ in the antiadiabatic regime, earlier we found that the CBM model undergoes a 
second-order quantum phase transition 
from a Luttinger liquid
to a CDW state at strong coupling \cite{au:rpys}. 
From Fig.~\ref{fig:cbmjtcomp},
 for both the CBM model and the CJT model, we see that the $S(\pi)$ curves coincide more or less. Hence,  up to half filling,
both the models exhibit similar 
CDW transition upon tuning electron-phonon coupling $g$.
However, above half filling in the CJT chain, particles will occupy the $d_{x^2-y^2}$ orbitals of the remaining sub-lattice 
because there is no repulsion between electrons in  the $d_{z^2}$ and $d_{x^2-y^2}$ orbitals on NN sites [see Eq. (\ref{eq:H_0})].
 Thus, in contrast to the CBM model, the particle-hole symmetry is broken for the CJT model!

In the subsequent sections, we shall study interacting CJT chains at couplings that are strong and weak.

\section{Interacting identically-long CJT chains at strong coupling}
\label{sec:strongG}
In the interacting CJT chains depicted in Fig.~\ref{fig:cartoon}$(c)$, 
we have C-type antiferromagnetism (i.e., each ferromagnetic chain is antiferromagnetically
coupled to the adjacent ones); consequently, there is no inter-chain hopping.
In these chains, hopping of electrons occurs only between the $d_{z^2}$ orbitals along the $z$-axis.
Furthermore, the chains are assumed  to have cooperative electron-phonon coupling  along the  $y$- and the $z$-axes, whereas,
along the x-axis the electron-phonon coupling is non-cooperative and is of the Holstein-type\cite{au:holstein,au:sdadys}.
Up to half filling, only the $d_{z^2}$ orbitals would be
occupied so that the system lowers its energy through the mobile electrons.
Here, we consider the interacting chains for only up to half filling; therefore, we neglect the  occupation of $d_{x^2-y^2}$ orbitals
in our analysis.
Hence, the lattice term of Eq.~\eqref{eq:gen_lattice} for the present case reduces to
\begin{IEEEeqnarray*}{rCl}
H^{ICJT}_l  = &&\> \frac{M}{2} \sum_{j,k} [\dot{u}^2_{x;0,j,k}+\dot{u}^2_{x;1,j,k} + \dot{u}^2_{y;j,k}
+ \dot{u}^2_{z;j,k}]\\
&&\>+\frac{K}{2} \sum_{j,k} [ u^2_{x;0,j,k}+u^2_{x;1,j,k}
 +u^2_{y;j,k} + u^2_{z;j,k} ].\IEEEyesnumber \IEEEeqnarraynumspace
\label{eq:intchainlatticeintermediate}
\end{IEEEeqnarray*}
We define $Q^{\prime}_{x;j,k}\equiv u_{x;1,j,k}+u_{x;0,j,k}$ and $Q_{x;j,k}\equiv u_{x;1,j,k}-u_{x;0,j,k}$
and recast Eq.~\eqref{eq:intchainlatticeintermediate} as follows:
\begin{IEEEeqnarray*}{rCl}
&&H^{ICJT}_l  \\
&&\quad= \frac{M}{2} \sum_{j,k} \left[\frac{1}{2}\left\{\dot{Q}^{\prime 2}_{x;j,k}+\dot{Q}^{2}_{x;j,k}\right\}
+\dot{u}^2_{y;j,k}+ \dot{u}^2_{z;j,k}\right]\\
&&\qquad+\frac{K}{2} \sum_{j,k} \left[\frac{1}{2} \left\{Q^{\prime 2}_{x;j,k}+Q^{2}_{x;j,k}\right\}
+u^2_{y;j,k}+u^2_{z;j,k} \right].\IEEEyesnumber \IEEEeqnarraynumspace 
\label{eq:intchainlattice}
\end{IEEEeqnarray*}
For the present system of interacting identically-long CJT chains, 
Eqs.~\eqref{eq:gen_hop_com} and \eqref{eq:gen_elph_com}
simplify to be
\begin{IEEEeqnarray}{rCl}
H^{ICJT}_t =-t \sum_{j,k} \left (d^\dagger_{z^2;j,k+1}d_{z^2;j,k}+ {\rm H.c.}\right ) ,
\label{eq:HICJT_t}
\end{IEEEeqnarray}
and 
\begin{IEEEeqnarray}{rCl}
\!\! H^{ICJT}_{ep} =
&-&g \omega_0 \sqrt{2 M \omega_0} \sum_{j,k}\Bigg [ \bigg \{(u_{z;j,k}-u_{z;j,k-1}) \nonumber \\
&+&\frac{1}{4}(Q_{x;j,k} 
+ u_{y;j,k}-u_{y;j-1,k})\bigg \} d^\dagger_{z^2;j,k}d_{z^2;j,k}\Bigg] .
\IEEEeqnarraynumspace
\label{eq:HICJT_ep}
\end{IEEEeqnarray}
Next, we observe that the displacement operator  $Q^{\prime }_{x;j,k}$ does not couple to the
electrons in the above Eqs. (\ref{eq:HICJT_t}) and (\ref{eq:HICJT_ep}); therefore, we drop terms
involving   $Q^{\prime }_{x;j,k}$ from Eq. (\ref{eq:intchainlattice}). Then, the total Hamiltonian
for the interaction CJT chains is given by 
\begin{IEEEeqnarray}{rCl}
 H^{ICJT} = H^{ICJT}_t+ H^{ICJT}_{ep}+ H^{ICJT}_l .
\label{eq:HICJT}
\end{IEEEeqnarray}
Now, representing the relevant displacement operators in second quantized 
form as
\begin{IEEEeqnarray*}{rCl}
u_{z;j,k} &=& \frac{a^\dagger_{z;j,k}+a_{z;j,k}}{\sqrt{2M\omega_0}},
~~~u_{y;j,k} = \frac{b^\dagger_{y;j,k}+b_{y;j,k}}{\sqrt{2M\omega_0}},\\
Q_{x;j,k}&=& \frac{c^\dagger_{x;j,k}+c_{x;j,k}}{\sqrt{2\frac{M}{2}\omega_0}}, 
\IEEEyesnumber
\label{eq:uzuyqx}
\end{IEEEeqnarray*}
and substituting in the total Hamiltonian of Eq. (\ref{eq:HICJT}), we obtain
\begin{IEEEeqnarray}{rCl}
\!\! H^{ICJT} =&-&t \sum_{j,k} (d^\dagger_{z^2;j,k+1}d_{z^2;j,k}+ {\rm H.c.}) \nonumber \\
&-&g \omega_0 \sum_{j,k}\bigg [(a^\dagger_{z;j,k}+a_{z;j,k})(n_{z^2;j,k}-n_{z^2;j,k+1}) \nonumber \\
&+&\frac{1}{4}(b^\dagger_{y;j,k}+b_{y;j,k})(n_{z^2;j,k}-n_{z^2;j+1,k}) \nonumber \\
&+&\frac{1}{2\sqrt{2}}(c^\dagger_{x;j,k}+c_{x;j,k})n_{z^2;j,k}\bigg ] \nonumber \\
&+& \omega_0 \sum_{j,k} (a^\dagger_{z;j,k}a_{z;j,k} + b^\dagger_{y;j,k}b_{y;j,k} + c^\dagger_{x;j,k}c_{x;j,k}), \nonumber \\
\IEEEeqnarraynumspace
\label{eq:intHam} 
\end{IEEEeqnarray}
where $n_{z^2;j,k}\equiv d^\dagger_{z^2;j,k}d_{z^2;j,k}$.
To obtain an effective Hamiltonian by performing perturbation theory,
we device a relevant Lang-Firsov transformation for the above Hamiltonian. 
In the polaronic frame of reference, $\tilde{H}^{ICJT} = \exp(S) H^{ICJT} \exp(-S)$
where the anti-Hermitian operator $S$ is identified as
\begin{IEEEeqnarray}{rCl}
S =&-&g \sum_{j,k}\bigg [(a^\dagger_{z;j,k}-a_{z;j,k})(n_{z^2;j,k}-n_{z^2;j,k+1}) \nonumber \\
&+&\frac{1}{4}(b^\dagger_{y;j,k}-b_{y;j,k})(n_{z^2;j,k}-n_{z^2;j+1,k}) \nonumber \\
&+&\frac{1}{2\sqrt{2}}(c^\dagger_{x;j,k}-c_{x;j,k})n_{z^2;j,k}\bigg ] .
\end{IEEEeqnarray}
Subsequently,  the Lang-Firsov transformed Hamiltonian is given by $\tilde{H}^{ICJT} = H_0^{ICJT} + H_1^{ICJT}$ with
\begin{IEEEeqnarray}{rCl}
\!\! H_0^{ICJT} =&&  \omega_0 \sum_{j,k} (a^\dagger_{z;j,k}a_{z;j,k} + b^\dagger_{y;j,k}b_{y;j,k} + c^\dagger_{x;j,k}c_{x;j,k}) \nonumber \\
&-& \frac{9}{4} g^2 \omega_0 \sum_{j,k} n_{z^2;j,k} \nonumber \\
&+& g^2 \omega_0 \sum_{j,k}\Big (2 n_{z^2;j,k}n_{z^2;j,k+1} 
+\frac{1}{8} n_{z^2;j,k}n_{z^2;j+1,k} \Big ) \nonumber \\
&-&te^{-\frac{13}{4} g^2} \sum_{j,k} (d^\dagger_{z^2;j,k+1}d_{z^2;j,k}+ {\rm H.c.}),
\end{IEEEeqnarray}
where terms $2g^2 \omega_0 \sum_{j,k} n_{z^2;j,k}n_{z^2;j,k+1}$ and
$ \frac{g^2 \omega_0}{8} \sum_{j,k} n_{z^2;j,k}n_{z^2;j+1,k}$
arise because of cooperative nature of the Jahn-Teller interaction along the $z$- and the $y$-directions, respectively.
Furthermore, the perturbation $H_1^{ICJT}$ is given by
\begin{IEEEeqnarray}{rCl}
H_1^{ICJT} 
&=& -t e^{-\frac{13}{4} g^2}
 \sum_{j,k} \bigg [ d^\dagger_{z^2;j,k+1}d_{z^2;j,k}
\{
{\cal{T}}_{+}^{j,k \dagger} {\cal{T}}^{j,k}_{-}
-1 \} \nonumber \\
&&\qquad \qquad + {\rm H.c.} \bigg ], 
\end{IEEEeqnarray}
where 
\begin{IEEEeqnarray}{rCl}
 {\cal{T}}^{j,k}_{\pm} 
& \equiv & \exp \Big [\pm g( 2 a_{z;j,k} - a_{z;j,k-1} - a_{z;j,k+1}) \nonumber \\
&& \pm \frac{g}{4}(b_{y;j-1,k+1} 
+b_{y;j,k}-b_{y;j-1,k}-b_{y;j,k+1}) \nonumber \\
&& \pm \frac{g}{2\sqrt{2}}(c_{x;j,k}-c_{x;j,k+1})\Big ].
\end{IEEEeqnarray}
Since we consider strong coupling and antiadiabaticity, we get the condition $t e^{-\frac{13}{4} g^2} << \omega_0$. 
Then, similar to the single chain CJT case [on identifying the eigenstates of $H_0^{ICJT}$ as $|n,m\rangle =|n\rangle_{el}\otimes |m\rangle_{ph}$
with $|0,0\rangle$ being the ground state], the second-order perturbation term
is given as:
\begin{IEEEeqnarray}{rCl}
H^{(2)} = \sum_{m}
\frac{\langle 0|_{ph} H_{1}^{ICJT} |m\rangle_{ph}
 \langle m|_{ph} H_{1}^{ICJT} |0\rangle_{ph}}
{E_{0}^{ph} - E_{m}^{ph}}.
\label{eq:H^2_int}
 \end{IEEEeqnarray}   
After tedious algebra, we get the the following expression for the effective Hamiltonian for interacting CJT chains:
\begin{widetext}
\begin{IEEEeqnarray*}{rCl}
 H^{ICJT}_{eff}
= &&-t e^{-\frac{13}{4} g^2}
 \sum_{j,k} (d^\dagger_{z^2;j,k+1}d_{z^2;j,k}+ {\rm H.c.}) \\
&&-\frac{t^2}{\omega_0}e^{-\frac{13}{2} g^2} 
G_5 \left(2,2,\frac{1}{8},\frac{1}{16},\frac{1}{16}\right)
\sum_{j,k} \left [d^\dagger_{z^2;j,k-1}(1-2n_{z^2;j,k})(1-n_{z^2;j-1,k})(1-n_{z^2;j+1,k})d_{z^2;j,k+1}+ {\rm H.c.} \right] \\
&&+2 \Bigg[ g^2 \omega_0 + \frac{t^2}{\omega_0}e^{-\frac{13}{2} g^2}
G_9 \left(4,1,1,\frac{1}{8},\frac{1}{8},\frac{1}{16},\frac{1}{16},\frac{1}{16},\frac{1}{16}\right)
 \Bigg]
\sum_{j,k} n_{z^2;j,k}n_{z^2;j,k+1}
+\frac{1}{8}g^2 \omega_0 \sum_{j,k} n_{z^2;j,k}n_{z^2;j+1,k} ,
\IEEEeqnarraynumspace
\label{eq:interchain_series}
\IEEEyesnumber
\end{IEEEeqnarray*}
where $G_n(\alpha_1,\alpha_2,\ldots,\alpha_n)$ are the same as those defined in the previous section. 
For  $g^2>>1$ (based on derivations in the Appendix \ref{app:serieses}), the above equation reduces to
\begin{IEEEeqnarray*}{rCl}
\frac{H^{ICJT}_{eff}}{t e^{-\frac{13}{4} g^2}}
= &-&\sum_{j,k} (d^\dagger_{z^2;j,k+1}d_{z^2;j,k}+ {\rm H.c.}) \\
&-&\frac{4}{17}\frac{te^{g^2}}{g^2\omega_0}\sum_{j,k} \left [d^\dagger_{z^2;j,k-1}(1-2n_{z^2;j,k})(1-n_{z^2;j-1,k})(1-n_{z^2;j+1,k})d_{z^2;j,k+1}
+{\rm H.c.}\right ] \\
&+&\left[ \frac{2g^2\omega_0}{t}  
+ \frac{4}{13} \frac{t}{g^2\omega_0}\right]e^{\frac{13}{4}g^2}\sum_{j,k} n_{z^2;j,k}n_{z^2;j,k+1}
+\frac{1}{8}\frac{g^2\omega_0}{t}e^{\frac{13}{4}g^2}\sum_{j,k} n_{z^2;j,k}n_{z^2;j+1,k} , 
\IEEEyesnumber
\label{eq:intchaineff}
\end{IEEEeqnarray*}
\end{widetext}
which is exactly the same as the single chain CJT result [see Eq.~\eqref{eq:1djteff}] 
except for the inter-chain repulsion
due to cooperative effects along the $y$-axis.
Furthermore, in the above equation, the occupancy-projection factors $(1-n_{z^2;j\pm1,k})$ [appearing in the NNN hopping term] 
arise due to the fact that the reduced hopping integral $t e^{-\frac{13}{4} g^2}$ is significantly smaller
than the inter-chain repulsion strength $\frac{g^2\omega_0}{8}$.

Although in the present work we  focus on  interacting chains in two-dimensions only,
effective Hamiltonian for interacting CJT chains in three-dimensions can be also obtained similarly.

Using the modified Lanczos algorithm of Ref. \onlinecite{au:gagliano}, in order to study 
the nature of the system
as the electron-phonon coupling $g$ is tuned, we diagonalize the Hamiltonian
in Eq.~\eqref{eq:interchain_series}
and calculate quantities such as density-density correlation function, structure factor, fidelity, and
fidelity susceptibility in the following subsections. Furthermore, a two-dimensional  system
with interacting CJT chains is taken to be mimicked by a two-chain CJT system with periodic
boundary conditions. Although we perform our calculations 
for a conservative value of $t/\omega_0=0.1$,  our results are valid for fairly larger values of
$t/\omega_0<1$ as demonstrated in the Appendix \ref{app:t_w_comp}.
\subsection{Density-density correlation function, structure factor, and order parameter}

\begin{figure}
\includegraphics[height=8cm,width=7.5cm,angle=-90]{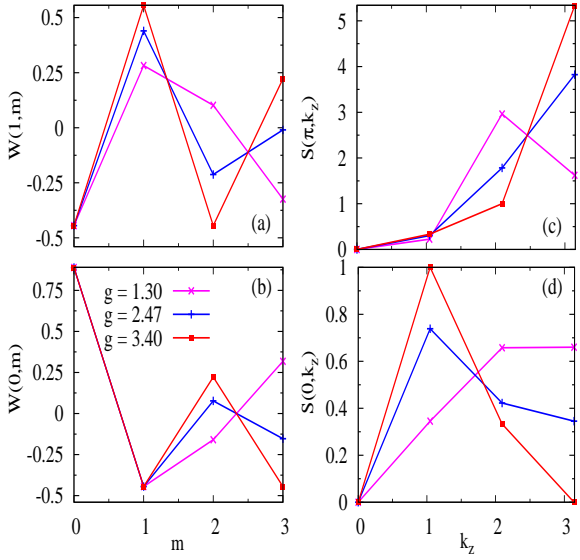} 
\caption{(Color online) Plots of (a) density-density inter-chain correlation function $W(1,m)$;
(b)  density-density intra-chain correlation function $W(0,m)$; 
(c) structure factor $S(\pi,k_z)$; and (d) structure factor $S(0,k_z)$
evaluated at adiabaticity $t/\omega_0 =0.1$, system size $N=12$ and particle number $ N_p = 4$.}
\label{fig:corr12_4}
\end{figure}

\begin{figure}
\includegraphics[height=8cm,width=7.5cm,angle=-90]{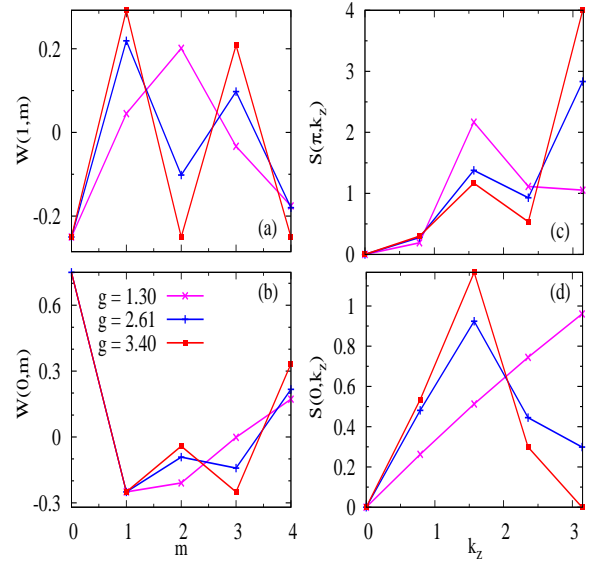} 
\caption{(Color online) Plots of (a) inter-chain correlation function $W(1,m)$;
(b)  intra-chain correlation function $W(0,m)$; 
(c) structure factor $S(\pi,k_z)$; and (d) structure factor $S(0,k_z)$
at  $t/\omega_0 =0.1$,  $N=16$ and  $ N_p = 4$.
}
\label{fig:corr16_4}
\end{figure}

\begin{figure}
\includegraphics[height=8cm,width=7.5cm,angle=-90]{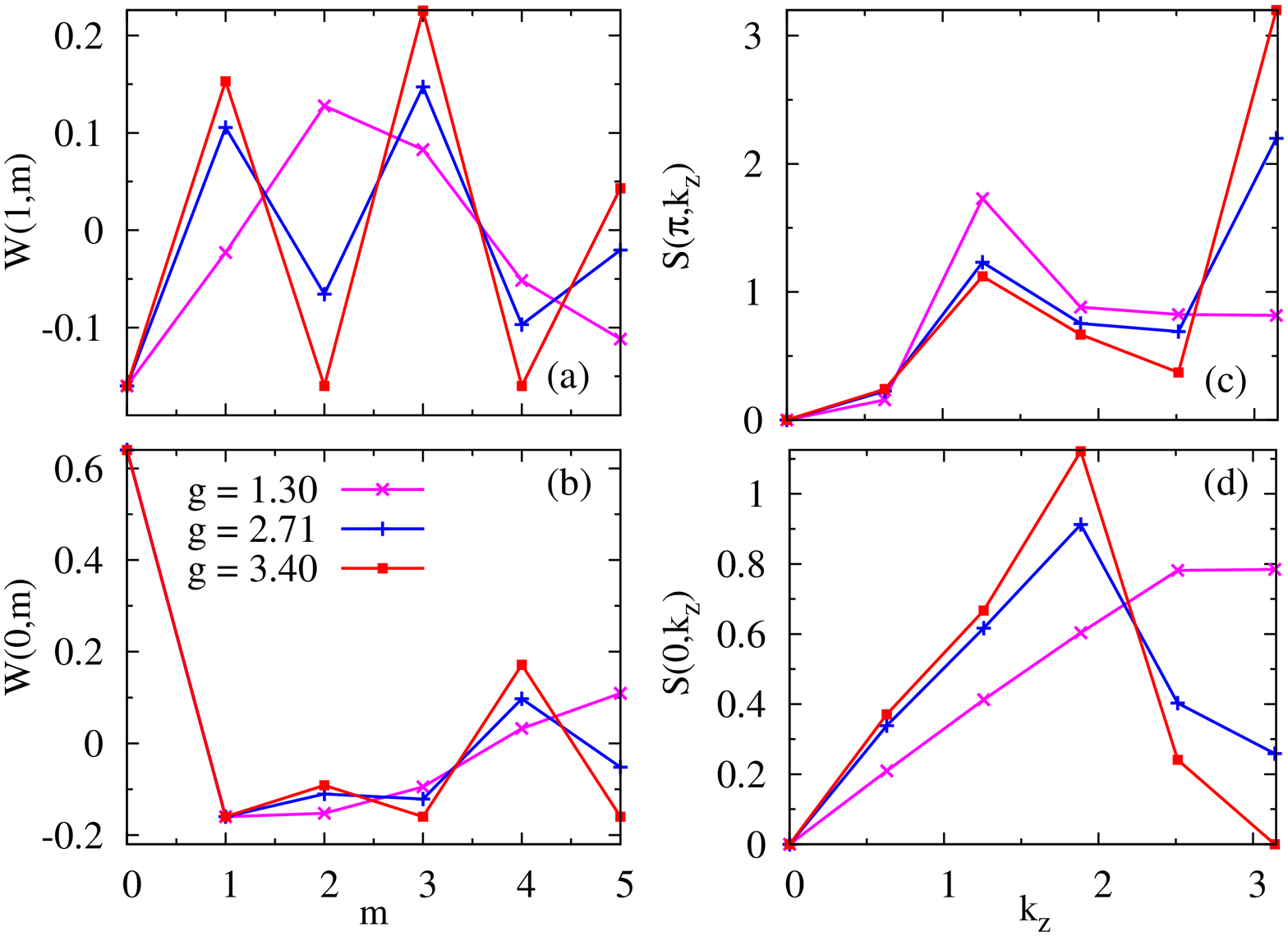} 
\caption{(Color online) Plots of (a) correlation function $W(1,m)$;
(b)  correlation function $W(0,m)$; 
(c) structure factor $S(\pi,k_z)$; and (d) structure factor $S(0,k_z)$
at  $t/\omega_0 =0.1$,  $N=20$ and  $ N_p = 4$.
}
\label{fig:corr20_4}
\end{figure}

\begin{figure}
\includegraphics[height=8cm,width=7.5cm,angle=-90]{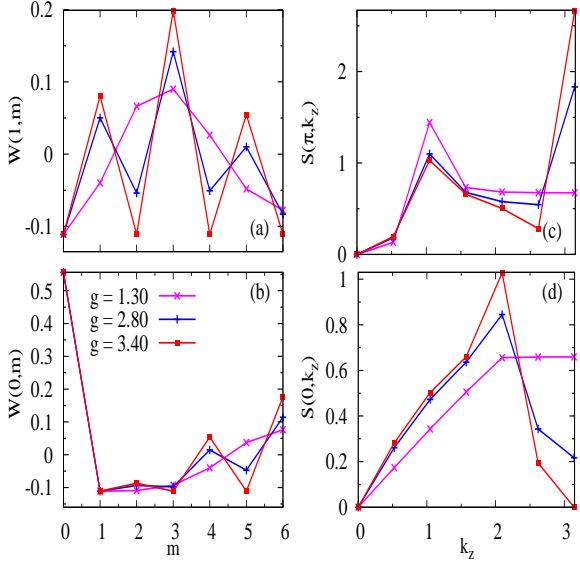} 
\caption{(Color online) Display of (a) two-point correlation function $W(1,m)$;
(b)  two-point correlation function $W(0,m)$; 
(c) structure factor $S(\pi,k_z)$; and (d) structure factor $S(0,k_z)$
at  $t/\omega_0 =0.1$,  $N=24$ and  $ N_p = 4$.
}
\label{fig:corr24_4}
\end{figure}

\begin{figure}
\includegraphics[height=8cm,width=7.5cm,angle=-90]{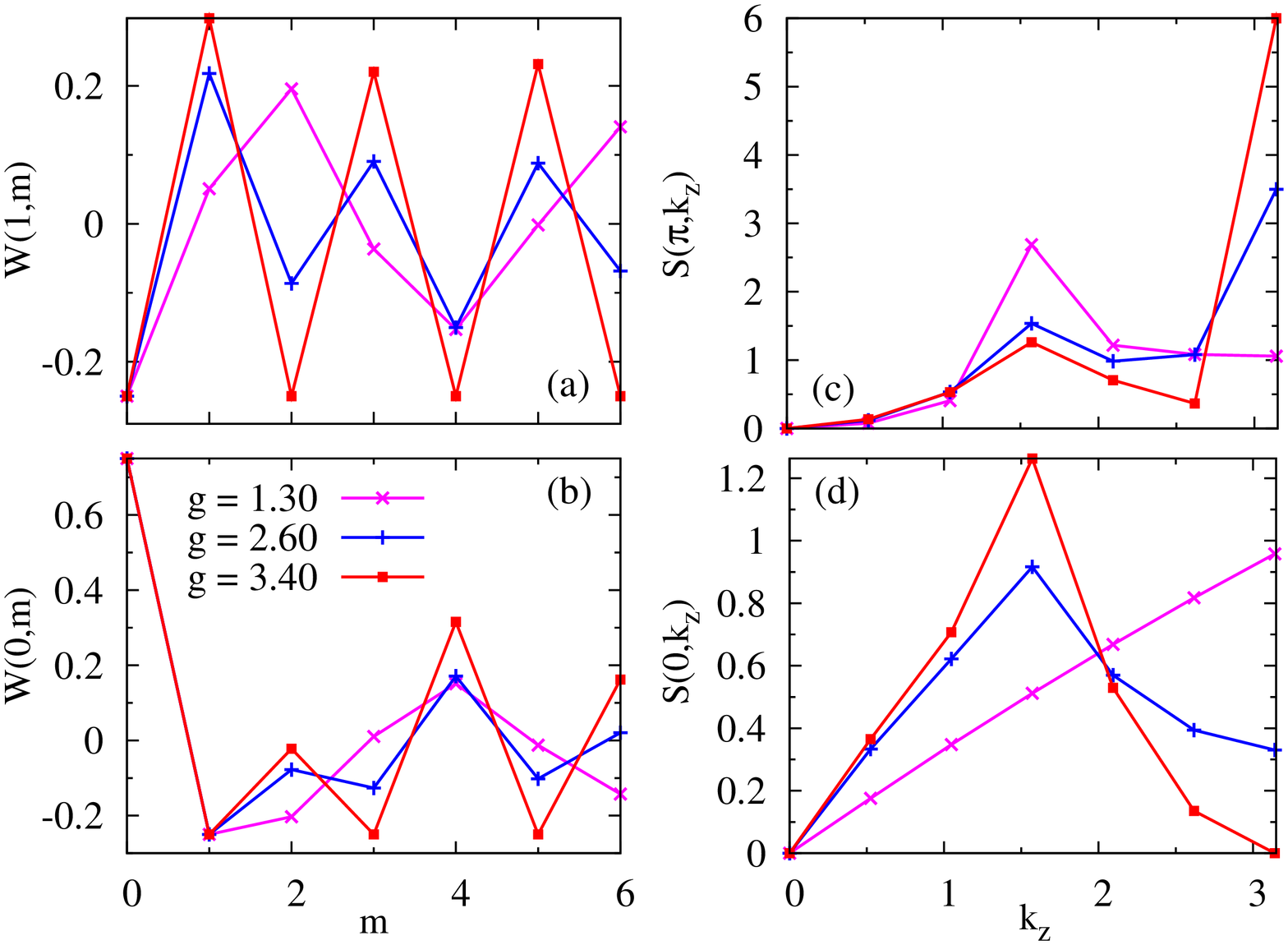} 
\caption{(Color online) Depiction of (a) correlation function $W(1,m)$;
(b)  correlation function $W(0,m)$; 
(c) structure factor $S(\pi,k_z)$; and (d) structure factor $S(0,k_z)$
at  $t/\omega_0 =0.1$,  $N=24$ and  $ N_p = 6$.}
\label{fig:corr24_6}
\end{figure}
\begin{figure}
\includegraphics[height=6.5cm,width=5.5cm,angle=-90]{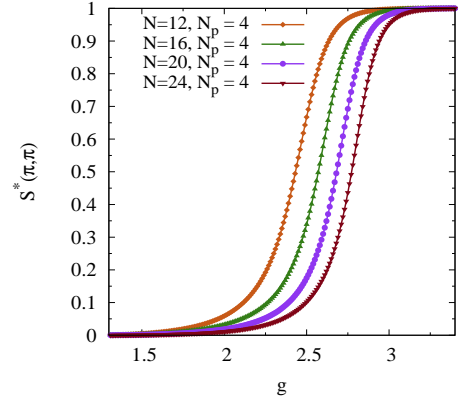} 
\caption{(Color online) Order parameter $S^*(\pi,\pi)$ 
as a function of the coupling $g$
when $t/\omega_0 = 0.1$.}
\label{fig:order_parameter}
\end{figure}

The two-point
correlation function, for density fluctuations of electrons that are apart by a
vector  $(l,m)$, is defined as 
\begin{IEEEeqnarray}{rCl}
W(l,m) &=& \frac{4}{N} \sum_{j,k}[\langle n_{z^2;j,k}n_{z^2;j+l,k+m}\rangle \IEEEnonumber \\
&&-\langle n_{z^2;j,k}\rangle \langle n_{z^2;j+l,k+m}\rangle ], 
\label{eq:corr}
\end{IEEEeqnarray}
for a  filling fraction $\nu=\langle n_{z^2;j,k}\rangle = \frac{N_p}{N}$, where $N$ is the total 
number of sites in the system and $N_p$  the total number of the particles in the system. The Fourier 
transform of $W(l,m)$, i.e., the structure factor $S(k_y,k_z)$, is given by
\begin{IEEEeqnarray}{c}
S(k_y,k_z) = \sum_{l,m} e^{i(k_yl+k_zm)} W(l,m), 
\end{IEEEeqnarray}
where for our two-chain CJT system, on assuming periodic boundary conditions, $k_y= \frac{2n_y\pi}{2} = n_y\pi$ with $n_y = 1,2$; and
$k_z= \frac{2n_z\pi}{\frac{N}{2}} = \frac{4n_z\pi}{N}$ with $n_z = 1,2,\ldots,\frac{N}{2}$.
We consider only even values of $N/2$ so as to obtain sublattice structure.
We will now derive exact relations for the structure factor
in a few special cases. To this end, the above equation is simplified to be
\begin{IEEEeqnarray}{rCl}
S\left(n_y\pi,\frac{4\pi n_z}{N}\right) = \sum_{m=1}^{N/2}\cos\left(\frac{4n_z\pi m}{N}\right)[(&&-1)^{n_y}W(1,m)\IEEEnonumber\\
&&+\>W(0,m)].
\IEEEeqnarraynumspace
\label{eq:S_gen} 
\end{IEEEeqnarray}
To begin with, we observe that
\begin{IEEEeqnarray*}{rCl}
S(\pi,\pi) = \left(\sum_{m_{\text{even}}}-\sum_{m_{\text{odd}}}\right)[-W(1,m)+W(0,m)] .
\end{IEEEeqnarray*}
Based on the derivation of  Eq.~\eqref{eq:app_spipi} in the Appendix \ref{app:str}, we can write 
\begin{IEEEeqnarray}{rCl}
S(\pi,\pi)= \frac{4}{N}\langle[(\hat{N}_{1,e}-\hat{N}_{1,o})+(\hat{N}_{2,o}-\hat{N}_{2,e})]^2\rangle, 
\label{eq:spipi}
\end{IEEEeqnarray}
where $\hat{N}_{j,e}=\sum_{k_{\text{even}}}n_{z^2;j,k}~ \left (\hat{N}_{j,o}=\sum_{k_{\text{odd}}}n_{z^2;j,k}\right )$
is the number operator for the total number of particles in the even (odd) sites of the $j$-{th} chain.
Next, again from Eq. (\ref{eq:S_gen}), we find 
\begin{IEEEeqnarray*}{rCl}
S(0,\pi) = \left(\sum_{m_{\text{even}}}-\sum_{m_{\text{odd}}}\right)[W(1,m)+W(2,m)] .
\end{IEEEeqnarray*}
As shown by the derivation of  Eq.~\eqref{eq:app_s0pi} in the Appendix \ref{app:str}, $S(0,\pi)$ can be expressed as
\begin{IEEEeqnarray}{rCl}
S(0,\pi)= \frac{4}{N}\langle[(\hat{N}_{1,e}-\hat{N}_{1,o})+(\hat{N}_{2,e}-\hat{N}_{2,o})]^2\rangle. 
\label{eq:s0pi}
\end{IEEEeqnarray}
Furthermore,  Eq. (\ref{eq:S_gen}) also yields
\begin{IEEEeqnarray*}{rCl}
S(\pi,0) = \left(\sum_{m_{\text{even}}}+\sum_{m_{\text{odd}}}\right)[-W(1,m)+W(2,m)].
\end{IEEEeqnarray*}
Then, from Eq.~\eqref{eq:app_spi0} in the Appendix \ref{app:str}, we observe that
\begin{IEEEeqnarray}{rCl}
S(\pi,0)= \frac{4}{N}\langle[(\hat{N}_{1,e}+\hat{N}_{1,o})-(\hat{N}_{2,e}+\hat{N}_{2,o})]^2\rangle . 
\label{eq:spi0}
\end{IEEEeqnarray}
Since hopping is permitted only along the chain, the particles will distribute themselves equally between the chains in order to
minimize the energy. 
Hence, 
for the full span of $g$
\begin{IEEEeqnarray}{rCl}
S(\pi,0)=0 ,
\label{eq:spi0_1} 
\end{IEEEeqnarray}
a fact verified by the depicted values of $S(\pi,0)$ in panel (c) of
 Figs. \ref{fig:corr12_4}, \ref{fig:corr16_4}, \ref{fig:corr20_4}, \ref{fig:corr24_4}, and \ref{fig:corr24_6}.
Additionally, we also have from Eq. (\ref{eq:S_gen})
\begin{IEEEeqnarray*}{rCl}
S(0,0) = \left(\sum_{m_{\text{even}}}+\sum_{m_{\text{odd}}}\right)[W(1,m)+W(2,m)].
\end{IEEEeqnarray*}
Then, from Eq.~\eqref{eq:app_s00} in the Appendix \ref{app:str}, we have
\begin{IEEEeqnarray}{rCl}
S(0,0)= \frac{4}{N}\langle[\hat{N}_p^2-N_p^2]\rangle, 
\label{eq:s00}
\end{IEEEeqnarray}
where $\hat{N}_p = (\hat{N}_{1,e}+\hat{N}_{1,o})+(\hat{N}_{2,e}+\hat{N}_{2,o})$ 
is the total particle number operator. Since we are working in a canonical ensemble 
and since the ground state is an eigenstate of  $\hat{N}_p$,  (at all couplings $g$) we get the simple relation
\begin{IEEEeqnarray}{rCl}
S(0,0)= 0 , 
\label{eq:s00_1}
\end{IEEEeqnarray}
which is manifested in panel (d) of 
 Figs. \ref{fig:corr12_4}, \ref{fig:corr16_4}, \ref{fig:corr20_4}, \ref{fig:corr24_4}, and \ref{fig:corr24_6}.

To understand the phase transition and the ordered state, we will now study some aspects of
the correlation function and the structure factor. We will consider
an extreme situation of the ordered state, namely, only one of the sub-lattices  of the two-chain system
is occupied. Then, in Eq. (\ref{eq:corr}),  for $l=1$ $(l=0)$ and $m$ being even (odd), we have $\langle n_{z^2;j,k}n_{z^2;j+l,k+m}\rangle = 0$.
Subsequently, we obtain
\begin{IEEEeqnarray}{rCl}
\label{eq:w0m_modd}
W(0,m_{\text{odd}}) = -\frac{4N^2_p}{N^2},
\end{IEEEeqnarray}
and
\begin{IEEEeqnarray}{rCl}
\label{eq:w1m_meven}
W(1,m_{\text{even}}) = -\frac{4N^2_p}{N^2}.
\end{IEEEeqnarray}
When only one of the two sub-lattices is occupied, the ground state becomes an eigenstate of the
operators $\hat{N}_{j,e}$, $\hat{N}_{j,o}$. Consequently, Eq.~\eqref{eq:spipi} and Eq.~\eqref{eq:s0pi} simply as follows:
\begin{IEEEeqnarray}{rCl}
[S(\pi,\pi)]_{\text{max}} &=& \frac{4N_p^2}{N}
\label{eq:spipimax} 
\end{IEEEeqnarray}
and
\begin{IEEEeqnarray}{rCl}
[S(0,\pi)]_{\text{min}} &=& 0.
\label{eq:s0pimin} 
\end{IEEEeqnarray}

Based on the above considerations, we will analyze the general features in the correlation functions
and structure factors at various filling fractions in the interacting CJT chains
by studying the  Figs.~\ref{fig:corr12_4}, \ref{fig:corr16_4}, \ref{fig:corr20_4},
 \ref{fig:corr24_4}, and \ref{fig:corr24_6}. In all these figures, we consider the system at three coupling strengths:
(a) when the system is quite disordered; (b) when the system is around the transition point; and (c) when the system is deep
inside the ordered phase.

First, panel (a) in all these figures 
shows the two-point inter-chain correlation function $W(1,m)$. 
 As the interaction strength is increased, the curves become more oscillatory;
at large coupling,  $W(1,m)$ attains the constant minimum value $-\frac{4N^2_p}{N^2}$
at all even values of $m$ as predicted by Eq. (\ref{eq:w1m_meven}) for the completely
ordered case. 
Second, panel (b) in all the Figs.  \ref{fig:corr12_4}--\ref{fig:corr24_6} portrays the
variation of the two-point intra-chain correlation function  $W(0,m)$. Here too,
the oscillations become more pronounced as the coupling strength $g$ increases.
Furthermore, at very strong coupling, 
$W(0,m)$ attains the same minimum value of $-\frac{4N^2_p}{N^2}$ [as given by Eq. (\ref{eq:w0m_modd})],
but at odd values of $m$ in contrast to $W(1,m)$.

Next, we observe that there are only two allowed values for the wavevector component $k_y$, i.e, $\pi$ and $0$.
We study the behavior of the structure factors $S(\pi,k_z)$ and $S(0,k_z)$ in the panels (c) and (d) of 
Figs.  \ref{fig:corr12_4}--\ref{fig:corr24_6}. In panel (c) of these figures,    at wave vector $(\pi,\pi)$,
$S(\pi,k_z)$ gradually attains its maximum value of $\frac{4N^2_p}{N}$ [in agreement with Eq. (\ref{eq:spipimax})]
as coupling becomes stronger. It is interesting to note that, in the disordered phase (i.e., at small coupling) 
$S(\pi,k_z)$ has a  peak at $k_z=2\pi\nu$ which is  reminiscent of the peaks in the Luttinger liquid 
for the CBM model (of Ref. \onlinecite{au:rpys}) and for the Holstein model (of Ref. \onlinecite{au:sdadys});
furthermore, the peak at $k_z=2\pi\nu$ becomes less pronounced as the  coupling $g$
increases in our interacting chains. This peak corresponds to only
short-range correlations; consequently, it is expected to be not observable in a thermodynamic system.
Last, in panel (d) of all the above-mentioned figures, $S(0,\pi)$ gradually decreases as the interaction strength
increases and finally drops to its lowest value of $0$ as estimated by Eq. (\ref{eq:s0pimin}).

We will now define the order parameter $S^*(\pi,\pi)$ as the following rescaled
value of $S(\pi,\pi)$:
\begin{IEEEeqnarray}{rCl}
S^*(\pi,\pi) \equiv \frac{S(\pi,\pi)-[S(\pi,\pi)]_\text{min}}{[S(\pi,\pi)]_\text{max}-[S(\pi,\pi)]_\text{min}} , 
\label{eq:orderparameter1}
\end{IEEEeqnarray}
where $[S(\pi,\pi)]_\text{min}$ 
and $[S(\pi,\pi)]_\text{max}$ are, 
respectively,  the minimum 
and the maximum 
values of $S(\pi,\pi)$.
 Hence, upon tuning electron-phonon coupling $g$ at any general filling up to half-filling, the order parameter $S^*(\pi,\pi)$ 
  varies from zero
 to one when the system transits from 
a charge disordered phase to a  CDW state [with ordering wavevector $(\pi,\pi)$]
as displayed in Fig. \ref{fig:order_parameter}. Interestingly, similar to the CBM model
of Ref. \onlinecite{au:rpys}, our interacting CJT model also
predicts a {\em conducting} commensurate  CDW state without an excitation gap .

\subsection{Ground-state fidelity and fidelity susceptibility} 
\begin{figure}
\includegraphics[height=8cm,width=10cm,angle=-90]{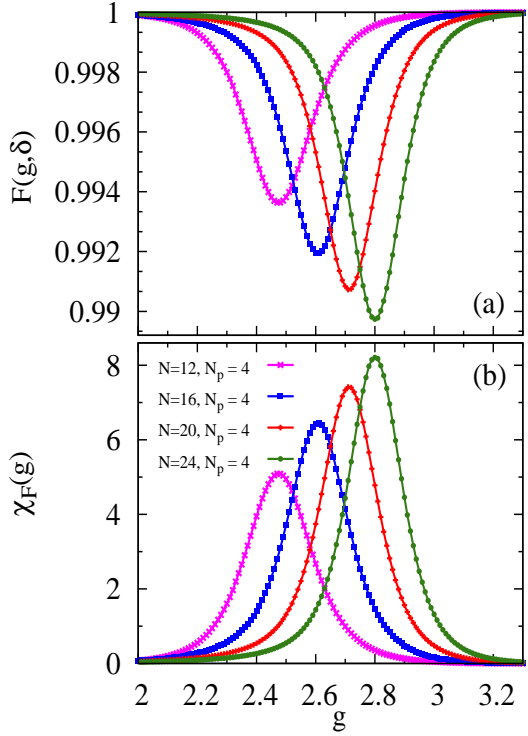}
\caption{(Color online) Plots of $(a)$ ground-state fidelity  and $(b)$ fidelity susceptibility 
evaluated at  $\delta = 0.05$ and at adiabaticity 
$t/\omega_0=0.1$.} 
\label{fig:fidelity}
\end{figure}

Although we analyzed the phase transition using the rescaled structure factor
$S^*(\pi,\pi)$ in the previous sub-section, to pinpoint the transition better we will use the ground-state fidelity and the fidelity susceptibility.
The ground-state fidelity, i.e., the overlap of the ground-state wavefunction at two very close values of 
the control parameter (i.e., at $g$ and $g+\delta$), is defined as follows \cite{au:zanardi}:
\begin{IEEEeqnarray}{c}
 F(g,\delta)=|\langle\Psi_0(g)|\Psi_0(g+\delta)\rangle| , 
\label{eq:fidelity}  
\end{IEEEeqnarray}
where $|\Psi_0 \rangle$ is the ground state of the system and $\delta$ is a small increment in $g$.
Next, fidelity susceptibility \cite{au:you} is defined as the second derivative of the ground-state fidelity:
\begin{IEEEeqnarray}{c}
 \chi_F(g) \equiv 
\partial^2_\delta F(g,\delta)|_{\delta=0}=2\lim_{\delta\rightarrow0}\frac{1-F(g,\delta)}{\delta^2} .
\label{eq:susceptibility}
\end{IEEEeqnarray}
It should be pointed out that, while the fidelity is dependent on the chosen value
of $\delta$, $\chi_F(g)$  is independent of it. For numerically evaluating
the values of fidelity and fidelity susceptibility, to avoid the problem due to degeneracy after phase transition,
we invoke the same method devised  in our earlier work \cite{au:rpys}. 

In Figs.~\ref{fig:fidelity}(a) and \ref{fig:fidelity}(b), we display our calculated ground-state fidelity and
the corresponding fidelity susceptibility 
for interacting-CJT-chain systems  of different sizes and at various values of strong-coupling.
The dips in the fidelity and
the concomitant peaks in the susceptibility are indicative of the location
of the phase transition in the $g$-parameter space; these transition points 
are in agreement with the order parameter curves in Fig.~\ref{fig:order_parameter}. Hence,
ground-state fidelity and fidelity susceptibility are useful tools for understanding phase transition in
our interacting CJT model. 

In the following section, we shall study interacting CJT chains
in the adiabatic regime and at weak coupling.
\section{Weak coupling analysis of interacting identically-long CJT chains}
\label{sec:weakG}

To analyze the weak-coupling case, 
we introduce new notation for ease of manipulation.  We adopt the notation that, for the $d_{z^2}$ orbital located at site $(j,k)$ in 
Fig.~\ref{fig:cartoon}$(c)$ and denoted by the destruction operator $d_{z^2;j,k}$,
 the surrounding oxygens in the y-direction are located at $u_{y;j-\frac{1}{2},k}$ and $u_{y;j+\frac{1}{2},k}$ 
while the adjacent oxygens in the z-direction are located at $u_{z;j,k-\frac{1}{2}}$ and $u_{z;j,k+\frac{1}{2}}$. 
Consequently, the corresponding second-quantized operators, analogous to those in Eq. (\ref{eq:uzuyqx}),
are given as 
\begin{IEEEeqnarray*}{rCl}
u_{z;j,k+\frac{1}{2}} &=& \frac{a^\dagger_{z;j,k+\frac{1}{2}}+a_{z;j,k+\frac{1}{2}}}{\sqrt{2M\omega_0}}, \\
u_{y;j+\frac{1}{2},k} &=& \frac{b^\dagger_{y;j+\frac{1}{2},k}+b_{y;j+\frac{1}{2},k}}{\sqrt{2M\omega_0}},\\
Q_{x;j,k}&=& \frac{c^\dagger_{x;j,k}+c_{x;j,k}}{\sqrt{2\frac{M}{2}\omega_0}}. 
\IEEEyesnumber
\label{eq:newuzuyqx}
\end{IEEEeqnarray*}
Then, for the interacting CJT chains, we rewrite  Eq.~\eqref{eq:intHam} as follows:
\begin{widetext}
\begin{IEEEeqnarray*}{rCl}
H^{ICJT} =&&-t \sum_{j,k} (d^\dagger_{z^2;j,k+1}d_{z^2;j,k}+ {\rm H.c.}) + \omega_0 \sum_{j,k} \big (a^\dagger_{z;j,k+\frac{1}{2}}a_{z;j,k+\frac{1}{2}} 
+ b^\dagger_{y;j+\frac{1}{2},k}b_{y;j+\frac{1}{2},k} + c^\dagger_{x;j,k}c_{x;j,k}\big ) \\
&&-g \omega_0 \sum_{j,k}\bigg [\Big \{\big (a^\dagger_{z;j,k+\frac{1}{2}}+a_{z;j,k+\frac{1}{2}}\big )
-\big (a^\dagger_{z;j,k-\frac{1}{2}}+a_{z;j,k-\frac{1}{2}}\big )\Big \}
+\frac{1}{4}\Big \{\big (b^\dagger_{y;j+\frac{1}{2},k}+b_{y;j+\frac{1}{2},k}\big )
-\big (b^\dagger_{y;j-\frac{1}{2},k}+b_{y;j-\frac{1}{2},k}\big )\Big \} \\ 
&& \qquad \qquad +\frac{1}{2\sqrt{2}}(c^\dagger_{x;j,k}+c_{x;j,k})\bigg ] n_{z^2;j,k} .
\IEEEyesnumber
\IEEEeqnarraynumspace
\end{IEEEeqnarray*} 
\end{widetext}
Next, we Fourier transform the above Hamiltonian using the following:
\begin{IEEEeqnarray*}{rCl}
d_{z^2;j,k} = \frac{1}{\sqrt{N}} \sum_{p,q} e^{i(pj+qk)} d_{z^2;p,q},
\end{IEEEeqnarray*} 
\begin{IEEEeqnarray*}{rCl}
 a_{z;j,k+\frac{1}{2}} = \frac{1}{\sqrt{N}} \sum_{p,q} e^{i\left[pj+q(k+\frac{1}{2})\right]} a_{z;p,q}, 
\end{IEEEeqnarray*}
\begin{IEEEeqnarray*}{rCl}
 b_{y;j+\frac{1}{2},k} = \frac{1}{\sqrt{N}} \sum_{p,q} e^{i\left[p(j+\frac{1}{2})+qk\right]} b_{y;p,q},
\end{IEEEeqnarray*}
and
\begin{IEEEeqnarray*}{rCl}
c_{x;j,k} = \frac{1}{\sqrt{N}} \sum_{p,q} e^{i(pj+qk)} c_{x;p,q} .
\end{IEEEeqnarray*}
We express $H^{ICJT}=H_0^{WC}+H_1^{WC}$ where
\begin{IEEEeqnarray}{rCl}
\!\!\! H_0^{WC} &=&-\sum_{p,q} \epsilon_{p,q} 
 d^\dagger_{z^2;p,q} d_{z^2;p,q} \nonumber \\
&& + \omega_0 \sum_{p,q} (a^\dagger_{z;p,q}a_{z;p,q}
+b^\dagger_{y;p,q}b_{y;p,q} +c^\dagger_{x;p,q}c_{x;p,q}), \nonumber \\
\IEEEeqnarraynumspace
\end{IEEEeqnarray}
with $\epsilon_{p,q}=-2t\cos(q)$
and 
\begin{IEEEeqnarray}{rCl}
H_1^{WC} = -\frac{g\omega_0}{\sqrt{N}} \sum_{p,q} Q_{eff}(p,q)
\rho_{p,q}, 
\end{IEEEeqnarray}
with $\rho_{p,q} = \sum_{{p^{\prime}},{q^{\prime}}} d^\dagger_{z^2;{p^{\prime}}+p,{q^{\prime}}+q} d_{z^2;{p^{\prime}},{q^{\prime}}} $
and 
\begin{IEEEeqnarray}{rCl}
Q_{eff}(p,q) &\equiv & 2i\sin\left(\frac{q}{2}\right)(a^\dagger_{z;-p,-q}+a_{z;p,q}) \nonumber \\
&&+\frac{i}{2}\sin\left(\frac{p}{2}\right)(b^\dagger_{y;-p,-q}+b_{y;p,q}) \nonumber \\
&&+\frac{1}{2\sqrt{2}}(c^\dagger_{x;-p,-q}+c_{x;p,q}) .
\end{IEEEeqnarray} 
We note that $Q_{eff}(p,q)$ is proportional to the Fourier transform of $\sqrt{2}Q_1+Q_3$ where $Q_1$ and $Q_3$ are the
breathing mode and a Jahn-Teller mode \cite{hotta}, respectively, at site $(j,k)$. 
The double time derivative of $Q_{eff}(p,q)$ is given by
\begin{IEEEeqnarray}{rCl}
\ddot{Q}_{eff}(p,q) = -\left[\left[Q_{eff}(p,q),H^{ ICJT}\right],H^{ ICJT}\right]. 
\label{eq:Qeff}
\end{IEEEeqnarray}
Let the eigenstates and eigenenergies of $H_0^{WC}$ be denoted by
$|\phi_l\rangle 
\equiv |n \rangle_{el} \otimes
|m_x,m_y,m_z\rangle_{ph}$ and $E^0_{\phi_l}$, respectively, with $ |\phi_0\rangle = |0;0_x,0_y,0_z \rangle$ 
being the ground state 
 with zero phonons. Furthermore, $\Phi_l$ and $E_{\Phi_l}$ are the corresponding eigenstates and eigenenergies, 
respectively, of  the interacting Hamiltonian $H^{ICJT}=H_0^{WC}+H_1^{WC}$. Then, the matrix elements of the above Eq.~(\ref{eq:Qeff}) can be
written as
\begin{IEEEeqnarray}{rCl}
\langle\Phi_l|\ddot{Q}_{eff}(p,q)|\Phi_{{l^{\prime}}}\rangle = -\left(E_{\Phi_l}-E_{\Phi_{{l^{\prime}}}}\right)^2
\langle\Phi_l|Q_{eff}(p,q)|\Phi_{{l^{\prime}}}\rangle . \nonumber \\ 
\end{IEEEeqnarray}
From the above equation, we see that during the transition from $|\Phi_{{l^{\prime}}}\rangle$ to $|\Phi_l\rangle$,
instability occurs if $\omega^2_{eff} \equiv \left(E_{\Phi_l}-E_{\Phi_{{l^{\prime}}}}\right)^2\leq 0$
provided that  $\langle\Phi_l|Q_{eff}(p,q)|\Phi_{{l^{\prime}}}\rangle \neq 0$.
In the weak-coupling regime, we have
\begin{IEEEeqnarray*}{rCl}
\omega^2_{eff}&=&\left(E_{\Phi_l}-E_{\Phi_{{l^{\prime}}}}\right)^2 
=\left\{\omega_0 + g^2 \omega_0^2 \> \text{Re}\> {\tilde{\chi}}_0(p,q;\omega_0)\right\}^2\\
&\approx&\omega_0^2 \left \{1 + 2g^2 \omega_0 \> \text{Re}\> {\tilde{\chi}}_0(p,q;\omega_0)\right \} ,
\end{IEEEeqnarray*}
where ${\tilde{\chi}}_0(p,q;\omega_0)$ is the relevant non-interacting dynamic polarizability defined below in Eq. (\ref{eq:chi0_tilde}).  
Therefore, the instability condition is given by
\begin{IEEEeqnarray}{rCl}
 1 + 2g^2 \omega_0 \> \text{Re}\> {\tilde{\chi}}_0(p,q;\omega_0)=0 .
\label{eq:instab}
\end{IEEEeqnarray}
For a detailed and rigorous derivation of this instability condition (involving the real part
of the non-interacting dynamic susceptibility), we refer the reader to Ref. \onlinecite{au:sdys}.

Now, up to second order in perturbation, the energy of the ground state of $H^{ICJT}$
is given by
\begin{IEEEeqnarray}{rCl}
E_{\Phi_0} &=& \langle \phi_0| H^{ICJT} |\phi_0 \rangle - \sum_{l \neq 0} \frac{\langle \phi_0| H_{1}^{WC} |\phi_l\rangle
 \langle \phi_l| {H^{WC}_{1}} |\phi_0\rangle}
{E^0_{\phi_l}-E^0_{\phi_0}} \nonumber \\
&=& E^0_{\phi_{0}} - \frac{g^2\omega^2_0}{N} \sum_{p,q,m\neq0}\bigg\{4\sin^2\left(\frac{q}{2}\right)
+\frac{1}{4}\sin^2\left(\frac{p}{2}\right)+\frac{1}{8}\bigg\} \nonumber \\
&&\>\times\frac{|\langle m|_{el} \rho_{p,q}|0\rangle_{el}|^2}{\xi_{m0}+\omega_0}, 
\label{eq:energy_phi}
\end{IEEEeqnarray}
where 
 $\xi_{m0}= \xi_m-\xi_0$ with $\xi_m$ being the energy of the state $|m\rangle_{el}$.

\begin{figure}
\includegraphics[height=9cm,width=7.5cm,angle=-90]{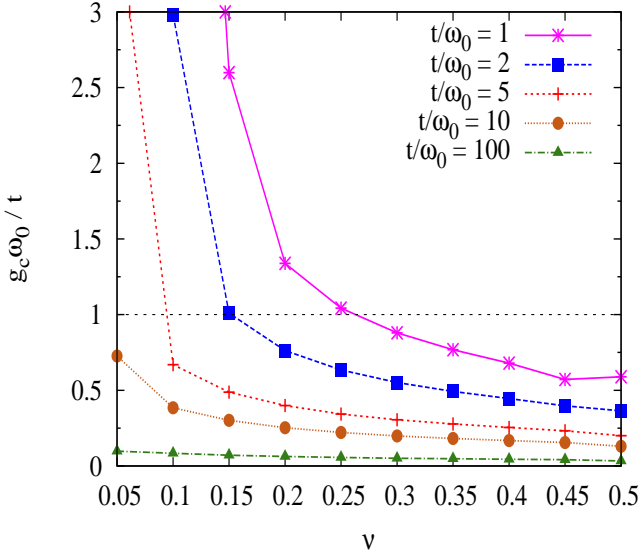}
\caption{(Color online) Plot of small parameter at critical coupling  (leading to density-dependent CDW transition)
for various filling fractions $\nu$ and adiabaticities $t/\omega_0$.} 
\label{fig:smallparameter}
\end{figure}

Next, we consider an excited state of $H_0^{IC}$ with a single phonon of momentum $(-p,-q)$
given by the general state
\begin{IEEEeqnarray}{rCl} 
|\psi^0_1\rangle &=& \beta \alpha |0;0_x,0_y,1_{z;-p,-q}\rangle +\beta\sqrt{1-\alpha^2}|0;0_x,1_{y;-p,-q},0_z\rangle \nonumber \\
&&+\sqrt{1-\beta^2}|0;1_{x;-p,-q},0_y,0_z\rangle , 
\end{IEEEeqnarray}
where $-1 \leq \alpha \leq 1$ and $-1 \leq \beta \leq 1$.  We evaluate the eigenenergy $E_{\Psi^0_1}$ for the interacting 
state $|\Psi^0_1\rangle$ of $H^{ICJT}$ as:
\begin{widetext}
\begin{IEEEeqnarray*}{rCl}
\!\! E_{\Psi^0_1}  &=& \langle \psi^0_1| H^{ICJT} |\psi^0_1 \rangle -\sum_{\phi_l\neq\psi^0_1} \frac{\langle \psi^0_1| H_{1}^{WC} |\phi_l\rangle
 \langle \phi_l| {H^{WC}_{1}} |\psi^0_1\rangle}
{E^0_{\phi_l}-E^0_{\psi^0_1}}\\ 
&=& E^0_{\phi_{0}} + \omega_0 - \frac{g^2\omega^2_0}{N}\sum_{p,q,m\neq0} \bigg\{4\sin^2\left(\frac{q}{2}\right)
+\frac{1}{4}\sin^2\left(\frac{p}{2}\right)+\frac{1}{8}\bigg\}
\frac{|\langle m|_{el} \rho_{p,q}|0\rangle_{el}|^2}{\xi_{m0}+\omega_0} \\
&&-\frac{g^2\omega^2_0}{N}\sum_{m\neq0} \Bigg[\bigg\{\bigg(2\beta\alpha\sin\left(\frac{q}{2}\right) 
 +\frac{1}{2}\beta\sqrt{1-\alpha^2}\sin\left(\frac{p}{2}\right)\bigg)^2+\frac{1}{8}(1-\beta^2)\bigg\}
\bigg\{\frac{|\langle m|_{el} \rho_{p,q}|0\rangle_{el}|^2}{\xi_{m0}+\omega_0}
+\frac{|\langle m|_{el} \rho_{-p,-q}|0\rangle_{el}|^2}{\xi_{m0}-\omega_0
}\bigg\}\Bigg]. \\
\IEEEyesnumber
\label{eq:energy_psi}
\end{IEEEeqnarray*}
After subtracting Eq.~\eqref{eq:energy_phi} from Eq.~\eqref{eq:energy_psi}, we obtain
\begin{IEEEeqnarray}{c}
E_{\Psi^0_1}- E_{\Phi_0} = \omega_0 + g^2 \omega^2_0 \> {\rm Re}\>{\tilde{\chi}}_0(-p,-q;\omega_0), 
\end{IEEEeqnarray}
where
\begin{IEEEeqnarray*}{rCl}
{\tilde{\chi}}_0(p,q;\omega_0)
\equiv \bigg\{\bigg(2\beta\alpha\sin\left(\frac{q}{2}\right) 
+\frac{1}{2}\beta\sqrt{1-\alpha^2}\sin\left(\frac{p}{2}\right)\bigg)^2
+\frac{1}{8}(1-\beta^2)\bigg\}
\chi_0(p,q;\omega_0) ,
\IEEEyesnumber
\label{eq:chi0_tilde}
\end{IEEEeqnarray*}
with 
\begin{IEEEeqnarray*}{rCl}
\chi_0(p,q;\omega_0) \equiv \frac{1}{N}\sum_{m\neq0} \Bigg[
\frac{|\langle m|_{el} \rho_{p,q}|0\rangle_{el}|^2}{\omega_0-\xi_{m0}+i\eta} 
-\frac{|\langle m|_{el} \rho_{-p,-q}|0\rangle_{el}|^2}{\omega_0+\xi_{m0}+i\eta}\Bigg]. 
\IEEEyesnumber
\label{eq:chi0}
\end{IEEEeqnarray*}
Based on the derivation in Appendix \ref{app:weakcoupling}, we have
\begin{IEEEeqnarray*}{rCl}
{\rm Re}\>\chi_0(p,q;\omega_0)
&=&\frac{1}{2\pi\omega_0}
\frac{\gamma}{\sqrt{1-\gamma^2}}
\Bigg\{\ln\Bigg[\frac{\big(1+\sqrt{1-\gamma^2}\big)^2-\big\{\gamma\tan\big(\frac{k_F}{2}+\frac{q}{4}\big)\big\}^2}
{\big(1-\sqrt{1-\gamma^2}\big)^2-\big\{\gamma\tan\big(\frac{k_F}{2}+\frac{q}{4}\big)\big\}^2}\Bigg]\\
&& \qquad \qquad \qquad \qquad -\ln\Bigg[\frac{\big(1+\sqrt{1-\gamma^2}\big)^2-\big\{\gamma\tan\big(-\frac{k_F}{2}+\frac{q}{4}\big)\big\}^2}
{\big(1-\sqrt{1-\gamma^2}\big)^2-\big\{\gamma\tan\big(-\frac{k_F}{2}+\frac{q}{4}\big)\big\}^2}\Bigg]\Bigg\} ,
\IEEEyesnumber
\end{IEEEeqnarray*}
where $\gamma \equiv \frac{\omega_0}{4t\sin(\frac{q}{2})} < 1$.
For values of the adiabaticity parameter $2 \lesssim t/\omega_0 \lesssim 5$ that 
occur in manganites \cite{tvr1} and for the range of filling fractions $0.05 \leq \nu \leq 0.5$ that include
 C-type antiferromagnetism in manganites, it should be pointed out that $\gamma < 1$.
In the above expression, it is interesting to note that $\chi_0(p,q;\omega_0)$ is independent of $p$.
To obtain the critical coupling $g_c$ at which the instability first sets in, we need to maximize
${\tilde{\chi}}_0(p,q;\omega_0)$ with respect to $p$.
It can be shown that the coefficient $\Big [\left\{2\beta\alpha\sin\left(\frac{q}{2}\right) 
+\frac{1}{2}\beta\sqrt{1-\alpha^2}\sin\left(\frac{p}{2}\right)\right\}^2+\frac{1}{8}(1-\beta^2)\Big ]$ 
[occurring in  Eq. (\ref{eq:chi0_tilde})] is maximized for $p=\pi$ with
 $\alpha = \frac{4\sin\left(\frac{q}{2}\right)}{\sqrt{1+16\sin^2\left(\frac{q}{2}\right)}}$ and
$\beta =1$. This value of $p=\pi$ is expected because of inter-chain particle repulsion. 
Therefore, the maximum  ${\tilde{\chi}}_0(p,q;\omega_0)$ is given by
\begin{IEEEeqnarray*}{rCl}
\left [{\rm Re}\>{\tilde{\chi}}_0(\pi,q;\omega_0)\right ]_{\rm max}
&=&\frac{1}{2\pi}\frac{1}{4\omega_0}\bigg\{1+16\sin^2\left(\frac{q}{2}\right)\bigg\}
\frac{\gamma}{\sqrt{1-\gamma^2}}
\Bigg\{\ln\Bigg[\frac{\big(1+\sqrt{1-\gamma^2}\big)^2-\big\{\gamma\tan\big(\frac{k_F}{2}+\frac{q}{4}\big)\big\}^2}
{\big(1-\sqrt{1-\gamma^2}\big)^2-\big\{\gamma\tan\big(\frac{k_F}{2}+\frac{q}{4}\big)\big\}^2}\Bigg]\\
&&\qquad \qquad \qquad -\ln\Bigg[\frac{\big(1+\sqrt{1-\gamma^2}\big)^2-\big\{\gamma\tan\big(-\frac{k_F}{2}+\frac{q}{4}\big)\big\}^2}
{\big(1-\sqrt{1-\gamma^2}\big)^2-\big\{\gamma\tan\big(-\frac{k_F}{2}+\frac{q}{4}\big)\big\}^2}\Bigg]\Bigg\} .
\IEEEyesnumber
\end{IEEEeqnarray*}
 Using the instability condition 
$1+2g^2_c\omega_0 \> [\text{Re}\> {\tilde{\chi}}_0(\pi,q;\omega_0)]_{\rm max} = 0 $ with $q=2k_F$ \cite{au:sdys},
 we get the critical coupling
$g_c$ for filling fraction $\nu=k_F/\pi$ as 
\begin{IEEEeqnarray}{rCl}
\frac{\pi}{g^2_c} &=& \frac{1}{4}\Big\{1+16\sin^2(k_F)\Big\}\frac{\gamma}{\sqrt{1-\gamma^2}} 
\Bigg\{\ln\Bigg[\frac{\big(1-\sqrt{1-\gamma^2}\big)^2-\big(\gamma\tan(k_F)\big)^2}
{\big(1+\sqrt{1-\gamma^2}\big)^2-\big(\gamma\tan(k_F)\big)^2}\Bigg]
-2\ln\left(\frac{1-\sqrt{1-\gamma^2}}{1+\sqrt{1-\gamma^2}}\right)\Bigg\}.
\label{eq:intchainweakgcritical} 
\end{IEEEeqnarray}
\end{widetext}
Based on the above Eq.~\eqref{eq:intchainweakgcritical}, Fig.~\ref{fig:smallparameter} displays the small parameter values at
the transition point $g=g_c$ for various fillings (up to half filling)
and for different values of the adiabaticity parameter. Our perturbation theory is valid
when the small parameter $g_c\omega_0/t<1$. 
 For the critical couplings $g_c$ in Fig.~\ref{fig:smallparameter}, the
system undergoes a quantum phase transition from a disordered state to an insulating CDW state with
ordering wave vector $(\pi,2\pi\nu)$.
As $t/\omega_0$ increases, even for  smaller filling fractions, phase transitions are predicted within 
our approach. 
Interestingly, for values of the adiabaticity parameter $2 \lesssim t/\omega_0 \lesssim 5$ that are relevant to manganites \cite{tvr1},
Fig.~\ref{fig:smallparameter} portrays the  critical coupling for density-dependent CDW ordering at filling fractions
where C-type antiferromagnetism is realized in manganites  \cite{au:chmaissem,au:szewczyk,maiti1,yao,au:kallias, au:goodenough}. 

\section{Discussion}
We will now present, at strong coupling, a simplification of the model for the interacting two-chain CJT system
 [represented by Eq. (\ref{eq:intchaineff})] by mapping
it onto  a model for a single CJT chain. For ease of discussion, for the two-chain case, we first recast Eq. (\ref{eq:intchaineff}) with
general coefficients: 
\begin{IEEEeqnarray*}{rCl}
H^{IC}_{eff} 
=&& -t_1 \sum_{j,k} (d^\dagger_{z^2;j,k+1}d_{z^2;j,k}+ {\rm H.c.}) \\
&&  -t_2\sum_{j,k} [d^\dagger_{z^2;j,k-1}(1-2n_{z^2;j,k}) \\
&&~~~~~~~~~~~~~ \times (1-n_{z^2;j+1,k})d_{z^2;j,k+1}
+{\rm H.c.}] \\
&&  +V_1\sum_{j,k} n_{z^2;j,k}n_{z^2;j,k+1} \\
&& + V_2\sum_{j,k} n_{z^2;j,k}n_{z^2;j+1,k}, 
\IEEEyesnumber
\label{new_intchain}
\end{IEEEeqnarray*}
where $V_1\rightarrow \infty$ and $V_2 \rightarrow \infty$.
The  two chains can be superposed on each other (to form a single new chain) with
the following conditions: (a) inter-chain particle repulsion $V_2 \rightarrow \infty$ 
implies that in the new chain there is also hard-core repulsion between the particles originating from different chains; (b)  to have more
number of unblocked sites for particles to hop to (so as to minimize the energy),
particles originating from different chains should alternate along the propagating z-axis
of the new chain;
and (c) because $V_1 \rightarrow \infty$, the next-nearest-neighbor hopping for particles originating
from  both the chains 
has the modified form $-t_2 d^{\dagger}_{z^2;k+1} (1-n_{z^2;k}) d_{z^2;k-1}$ where $d^{\dagger}_{z^2;k}$ is the creation operator
of a particle at site $k$ of the new chain and $n_{z^2;k} =d^{\dagger}_{z^2;k} d_{z^2;k}$. On incorporating
the above conditions,  the $H^{IC}_{eff}$ of Eq. (\ref{new_intchain}) has a simplified
mapping onto the following Hamiltonian of a single chain:
\begin{IEEEeqnarray*}{rCl}
H^{SC}_{eff} 
&=& -t_1 \sum_{k} (d^\dagger_{z^2;k+1}d_{z^2;k}+ {\rm H.c.}) \\
&&  -t_2\sum_{k} [d^\dagger_{z^2;k+1}(1-n_{z^2;k})d_{z^2;k-1}
+{\rm H.c.}]  .
\IEEEyesnumber
\IEEEeqnarraynumspace
\label{new_sing_chain}
\end{IEEEeqnarray*}
Thus a system of $N$ sites and $N_p$ particles in the interacting-chain model
(given by $H^{IC}_{eff}$) has been reduced to a system of $N/2$ sites with $N_p$ particles
[and governed by Eq. (\ref{new_sing_chain})],
thereby leading to a significant increase in the size of computationally accessible systems;
the number of basis states in the explored Hilbert space reduces from ${\rm {^NC_{N_p}}}$ to ${\rm {^{\frac{N}{2}}C_{N_p}}}$.
Furthermore, the Hamiltonian for a single CJT chain as given by Eq. (\ref{eq:1djteff}), for a system of $N$ sites and $N_p$ particles,
can be reexpressed in a generalized form
\begin{IEEEeqnarray*}{rCl}
 H^{C}_{eff} 
&=&-t_1 \sum_k (d^\dagger_{z^2;k+1}d_{z^2;k}+ {\rm H.c.} )\\
&&-t_2\sum_k \left[d^\dagger_{z^2;k-1}(1-2n_{z^2;k})d_{z^2;k+1}
+ {\rm H.c.}\right] \\
&&+ V \sum_k n_{z^2;k} n_{z^2;k+1} ,
\IEEEyesnumber
\IEEEeqnarraynumspace
\label{new_1djteff}
\end{IEEEeqnarray*} 
with $V \rightarrow \infty$.
Then, as pointed out in Ref. \onlinecite{au:rpys}, the model of the above Eq. (\ref{new_1djteff})
can be mapped onto the following
model with $N-N_p$ sites and $N_p$ particles:
\begin{IEEEeqnarray*}{rCl}
&& H^{RC}_{eff} \\
=&&-t_1 \sum_k (d^\dagger_{z^2;k+1}d_{z^2;k}+ {\rm H.c.} )\\
&&-t_2\sum_k \left[d^\dagger_{z^2;k-1}(1-n_{z^2;k})d_{z^2;k+1}
+ {\rm H.c.}\right] .
\IEEEyesnumber
\label{new2_1djteff}
\end{IEEEeqnarray*} 
Here, it should be noted that the Hamiltonians in Eqs. (\ref{new_sing_chain}) and (\ref{new2_1djteff})
are identical.
From the above considerations, it follows that  the evolution of the order parameter, fidelity,
and fidelity susceptibility in the interacting two-chain system 
with $N=n$ sites and $N_p=n_p$ particles [and represented by Eq. (\ref{eq:intchaineff})] is identically mimicked by a single chain system
with $N=n_p+n/2$ sites and $N_p=n_p$ particles [and represented by Eq. (\ref{eq:1djteff})] as shown in Figs. \ref{op} and \ref{map_fid}
for $n=16$ and $n_p=4$. Additionally, we also note that the nature of the phase transition for the CBM model of Eq. (\ref{eq:1dcbmeff})
[and studied in Ref. \onlinecite{au:rpys}] is the same as that for the one-dimensional CJT system represented
by Eq. (\ref{eq:1djteff}) because of the close similarity of the two governing equations and the structure factors depicted
in Fig. {\ref{fig:cbmjtcomp}. Furthermore, since the CBM model of Ref. \onlinecite{au:rpys} undergoes a second-order
phase transition and because the interacting-chain JT system can be mapped onto the single-chain JT system, we expect
the phase transition (that occurs when the coupling $g$ is varied) in interacting CJT chains to be also  second-order.

\begin{figure}
\includegraphics[height=8cm,width=5.5cm,angle=-90]{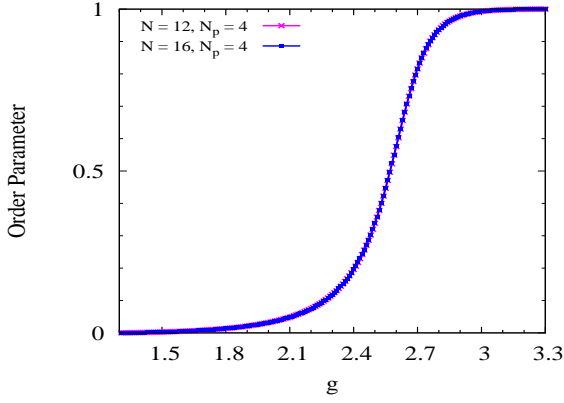}
\caption{(Color online) Plots of the order parameters verifying that
a system of interacting CJT chains with $N = 16$ sites
and $N_p=4$ particles is  closely mimicked by a single CJT chain with
$N = 12$ sites and $N_p =4$ particles.
} 
\label{op}
\end{figure}
 
\begin{figure}
\includegraphics[height=6cm,width=8.0cm,angle=-90]{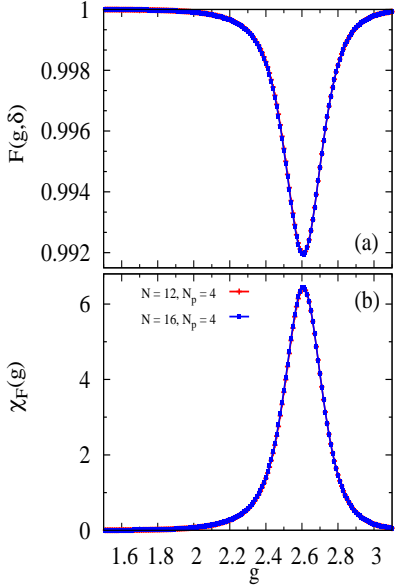}
\caption{(Color online) Plots of (a) ground-state fidelity and (b) fidelity susceptibility
showing very similar phase transition for a system of interacting CJT chains (with $N = 16$ sites
and $N_p=4$ particles) and a system of a single CJT chain (with
$N = 12$ sites and $N_p =4$ particles).} 
\label{map_fid}
\end{figure}

Next,  we would like to present the relevance  of our derived model (for interacting CJT chains
in the strong coupling regime) to a different system. 
A wide variety of condensed matter models can be simulated by suitable engineering of ultra-cold atom systems
in optical lattices.
We would like to point out that our studied model of interacting
chains can be realized in a cold-atom system represented by hard-core-bosons (HCBs). 
To this end, we transform a fermion (represented by a creation operator $d_{z^2}^{\dagger}$)
 to a HCB (represented by a creation operator $h^{\dagger}$) by using the relation
$h^{\dagger}_{j,k} =d^{\dagger}_{z^2;j,k} \Pi_{i<j} (1-2n_{z^2;i,k})$;
this transformation is equivalent to a Wigner-Jordan transformation followed by a spin-to-HCB transformation.
Then the equivalent of the fermionic model in Eq. (\ref{eq:intchaineff}) is given by the following model for HCBs:  
\begin{IEEEeqnarray*}{rCl}
&& \!\!\!\!\!\!\!\!\!\! {H^{2D}_{\rm HCB}} \\
&= &-t_1 \sum_{j,k} (h^\dagger_{j,k}h_{j,k+1}+ {\rm H.c.}) \\
&&- t_2 \sum_{j,k} [h^\dagger_{j,k-1}(1-n^h_{j-1,k})(1-n^h_{j+1,k})h_{j,k+1}
+{\rm H.c.}] \\
&& + V_1\sum_{j,k} n^h_{j,k}n^h_{j,k+1}
+ V_2\sum_{j,k} n^h_{j,k}n^h_{j+1,k}, 
\IEEEyesnumber
\label{2D_HCB}
\end{IEEEeqnarray*}
where $n^h_{j,k}$ ($\equiv h^{\dagger}_{j,k}h_{j,k}$) is the number operator for HCBs.
As pointed out in Refs. \onlinecite{becca,struck}, nearest-neighbor and next-nearest-neighbor hopping terms
can be tuned independently; additionally, nearest-neighbor repulsion can also be achieved \cite{cirac}.
 Thus, ladders and two-dimensional systems in general [which realize the Hamiltonian in Eq. (\ref{2D_HCB})]
can be simulated experimentally for a broad range of values of the parameters $t_1$, $t_2$, $V_1$, and $V_2$ . It is interesting to note that a 
one-dimensional version of the above Hamiltonian, i.e,
\begin{eqnarray}
 {H^{1D}_{\rm HCB}}
= &-&t_1 \sum_{k} (h^\dagger_{k}h_{k+1}+ {\rm H.c.})
- t_2 \sum_{k} (h^\dagger_{k-1}h_{k+1} + {\rm H.c.}) \nonumber \\
&+&  V\sum_{k} n^h_{k}n^h_{k+1}.
\label{t1t2V}
\end{eqnarray}
 was studied  by Mishra et al. \cite{mishra1,mishra2}.
These authors considered kinetic frustration and obtained the phase diagram.
Furthermore, in the extreme case of $t_1=0$, the above model [of Eq. \eqref{t1t2V}]
was recently shown to undergo a striking discontinuous transition from a superfluid
to a supersolid in Ref. \onlinecite{ag_sy}.

\section{Conclusions}
In this paper, we studied the CDW transition in interacting CJT chains at strong coupling [characterized by \cite{au:rpys} $t/(g\omega_0) < 1$]
and at weak coupling [identified by \cite{au:sdys} $(g\omega_0)/t < 1$]. The nature of CDW in the interacting CJT chains
is quite different in the two cases:
strong coupling analysis predicts a conducting  CDW with $Z_2$ symmetry being broken (i.e., one sub-lattice
has higher density), whereas weak coupling theory
shows that an insulating  CDW results with a wavevector that varies linearly with density. Our study is relevant
in manganite systems 
 with C-type antiferromagnetism and suggests identifying the charge-ordering wavevector
as an alternate way
for determining the controversial regime of the electron-phonon coupling. 
In the case where the coupling is strong, we use structure factor and fidelity to track the
quantum phase transition. Exact expressions for the structure factor  at special
wavevectors [i.e., $(0,0)$, $(\pi,0)$, $(0,\pi)$, and $(\pi,\pi)$] are derived
to understand the evolution of the system when the tuning parameter is varied. 
On the other hand, in the weak coupling regime, we identify the critical coupling
at which the phonon mode (involving the breathing mode and a Jahn-Teller mode) becomes soft; our analysis is applicable for 
the adiabaticities occurring in manganites.
Although our analysis for the interacting CJT chains corresponds to a two-dimensional case, it
 is extendable to three-dimensional systems as well and the conclusions
are expected to be similar.

\section{Acknowledgments}
We thank P. Majumdar, P. Littlewood, N. D. Mathur, D. E. Khmelnitskii, and A. Ghosh for useful discussions.
\appendix
\section{Simplification of  the function $G_n(\alpha_1,\alpha_2,\ldots,\alpha_n)$}
\label{app:serieses}
In this appendix, we obtain simple expressions for the function $G_n(\alpha_1,\alpha_2,\ldots,\alpha_n)$
appearing in the main text.
The general term  $G_n(\alpha_1,\alpha_2,\ldots,\alpha_n)$ 
is defined as 
\begin{IEEEeqnarray*}{rCl}
G_n(\alpha_1,\alpha_2,\ldots,\alpha_n) &\equiv& F_n(\alpha_1,\alpha_2,\ldots,\alpha_n)\\
&&+\sum_{k=1}^{n-1}\sum_c
F_k(\alpha_{c_1},\alpha_{c_2},\ldots,\alpha_{c_k}) , 
 \end{IEEEeqnarray*}   
where 
\begin{IEEEeqnarray*}{rCl}
\!\!\!\! F_n(\alpha_1, \ldots , \alpha_n ) \equiv \sum_{m_1=1}^{\infty} 
 ...
\sum_{m_n=1}^{\infty}
 \frac {(\alpha_1 g^2)^{m_1} \ldots (\alpha_n g^2)^{m_n}}
{m_1!\ldots m_n!(m_1+ \ldots + m_n)} ,
 \end{IEEEeqnarray*}   
and the summation over $c$ represents summing over all possible
$^nC_m$ combinations of $m$ arguments chosen from the total set of 
 $n$ arguments $\{\alpha_1,\alpha_2,\ldots,\alpha_n\}$.

We begin by examining the simple case of the term $G_2(2,2)$ $[=F_2(2,2)+2F_1(2)]$ appearing
in Eq. \eqref{eq:1dcbmseries}. We evaluate the derivative, with respect to $g^2$,  of 
$G_2(2,2)$ as follows:
\begin{IEEEeqnarray*}{rCl}
g^2\frac{d}{dg^2}G_2(2,2)
&&=(e^{2g^2}-1)(e^{2g^2}-1)+2(e^{2g^2}-1) \\
&&=\{(e^{2g^2}-1)+1\}\{(e^{2g^2}-1)+1\}-1\\
&&= e^{4g^2}-1. \IEEEyesnumber
\end{IEEEeqnarray*}
Then, on performing integration, we get
\begin{IEEEeqnarray*}{rCl}
G_2(2,2)
&=&\int \frac{e^{4g^2}-1}{g^2}dg^2 \\
&=& \int \sum^\infty_{n=1} \frac{4^n(g^2)^{(n-1)}}{n!}dg^2\\
&=& \sum^\infty_{n=1} \frac{(4g^2)^n}{n\>n!}.
\IEEEyesnumber
\end{IEEEeqnarray*}
Next, we consider 
the other term
$G_3(4,1,1)$ $[=F_3(4,1,1)+2F_2(4,1)+F_2(1,1)+F_1(4)+2F_1(1)]$
occurring in Eq.~\eqref{eq:1dcbmseries}. On taking the derivative of $G_3(4,1,1)$ with respect to $g^2$, we obtain
\begin{IEEEeqnarray*}{rCl}
&&g^2\frac{d}{dg^2}G_3(4,1,1)
\\
&&\quad =(e^{4g^2}-1)(e^{g^2}-1)(e^{g^2}-1)
+2(e^{4g^2}-1)(e^{g^2}-1)\\
&&\qquad +(e^{g^2}-1)(e^{g^2}-1)+(e^{4g^2}-1)+2(e^{g^2}-1)\\
&&\quad=\{(e^{4g^2}-1)+1\}\{(e^{g^2}-1)+1\}\{(e^{g^2}-1)+1\}-1\\
&&\quad 
= e^{6g^2}-1. \IEEEyesnumber
\end{IEEEeqnarray*}
Hence, we get
\begin{IEEEeqnarray*}{rCl}
G_3(4,1,1)
&=&\int \frac{e^{6g^2}-1}{g^2}dg^2 \\
&=& \int \sum^\infty_{n=1} \frac{6^n(g^2)^{(n-1)}}{n!}dg^2\\
&=& \sum^\infty_{n=1} \frac{(6g^2)^n}{n\>n!}.\IEEEyesnumber
\end{IEEEeqnarray*}

Finally, we evaluate the derivative of the general term
$G_n(\alpha_1,\alpha_2,\ldots,\alpha_n)$
with respect to  $g^2$:
\begin{IEEEeqnarray*}{rCl}
&&g^2\frac{d}{dg^2}G_n(\alpha_1,\alpha_2,\ldots,\alpha_n)
\\
&&\quad=(e^{\alpha_1g^2}-1)(e^{\alpha_2g^2}-1)\ldots(e^{\alpha_ng^2}-1)\\
&&\qquad +\sum_{k=1}^{n-1}\sum_c (e^{\alpha_{c_1}g^2}-1)(e^{\alpha_{c_2}g^2}-1)\ldots(e^{\alpha_{c_k}g^2}-1)\\
&&\quad=\left [ \Pi_{i=1}^{n}\big \{(e^{\alpha_ig^2}-1)+1\big \}\right ]-1\\
&&\quad =e^{\sum_{i=1}^n\alpha_ig^2}-1. \IEEEyesnumber
\end{IEEEeqnarray*}
Then, the general term is obtained to be 
\begin{IEEEeqnarray*}{rCl}
G_n(\alpha_1,\alpha_2,\ldots,\alpha_n)
&=&\int \frac{e^{\sum_{i=1}^n\alpha_ig^2}-1}{g^2}dg^2\\
&=& \int \sum^\infty_{m=1} \frac{(\sum_{i=1}^n\alpha_i)^m(g^2)^{(m-1)}}{m!}dg^2\\
&=& \sum^\infty_{m=1} \frac{(\sum_{i=1}^n\alpha_ig^2)^m}{m\>m!}.\IEEEyesnumber
\end{IEEEeqnarray*}
For large values of $g^2$, we have the approximation
\begin{IEEEeqnarray*}{rCl}
\int \frac{e^{\sum_{i=1}^n\alpha_ig^2}-1}{g^2}dg^2
\approx\frac{e^{\sum_{i=1}^n\alpha_ig^2}}{\sum_{i=1}^n\alpha_ig^2}.
\end{IEEEeqnarray*}
\section{Comparison of phase transition in interacting CJT chains at strong coupling and various values of adiabaticity}
\label{app:t_w_comp}
\begin{figure}[t]
\includegraphics[height=8.5cm,width=7.5cm,angle=-90]{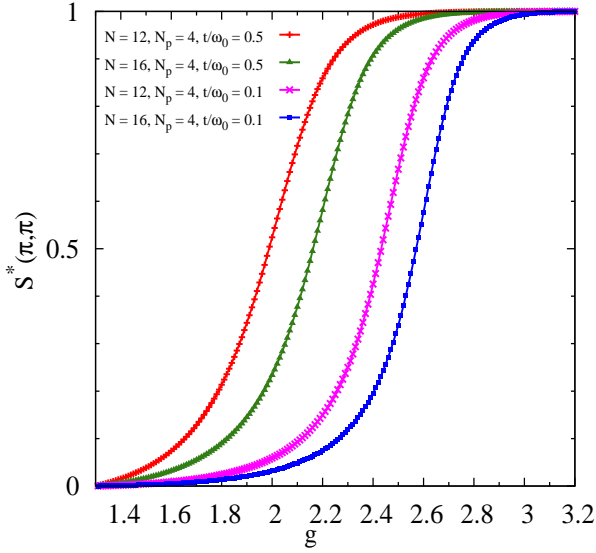}
\caption{(Color online) Plots comparing order parameter $S^*(\pi,\pi)$ at different values of $t/\omega_0$
and at different filling fractions.} 
\label{fig:app_order_tw}
\end{figure}
In this appendix, we compare the  behavior of the interacting CJT chains at strong coupling
and for different values of the adiabaticity parameter $t/\omega_0$. In Fig.~\ref{fig:app_order_tw}
we portray the order parameter $S^*(\pi,\pi)$ for two filling fractions and for two sufficiently different values of the adiabaticity
parameter. As is evident from the figure,  for both the fillings, the transition for $t/\omega_0=0.5$ takes place at smaller values of $g$
 compared to the transitions for  $t/\omega_0=0.1$.
This observation is consistent with the transitions indicated by the extrema in the plots 
of ground-state fidelity and  fidelity susceptibility displayed 
in Fig.~\ref{fig:app_fid_tw}. 
\begin{figure}[t]
\includegraphics[height=9cm,width=9cm,angle=-90]{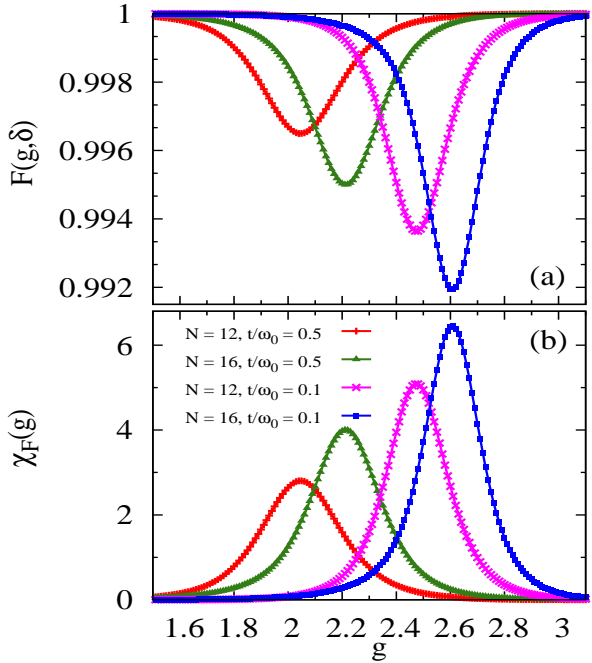}
\caption{(Color online) Plots comparing $(a)$ ground-state fidelity  and $(b)$ fidelity susceptibility
at  $\delta=0.05$ and $N_p=4$ when using different values of $t/\omega_0$ and fillings $N_p/N$.} 
\label{fig:app_fid_tw}
\end{figure}
\section{Exact expressions for the structure factor in limiting cases}
\label{app:str}
In this appendix, we derive formulae for  $S(\pi,\pi)$, $S(0,\pi)$, $S(\pi,0)$, and $S(0,0)$. 
The structure factor is expressed in terms of the correlation function $W(l,m)$ as follows:
\begin{IEEEeqnarray}{c}
S(k_y,k_z) = \sum_{l,m} e^{i(k_yl+k_zm)} W(l,m), 
\end{IEEEeqnarray}
where $k_y = n_y\pi$ with $n_y = 1, 2$ and
$k_z=
 \frac{4n_z\pi}{N}$ with $n_z = 1,2,\ldots,
\frac{N}{2}$.
Since we are dealing with two interacting chains, 
the above equation simplifies to
\begin{IEEEeqnarray}{rCl}
S\left(n_y\pi,\frac{4\pi n_z}{N}\right) = \sum_{m=1}^{N/2}\cos\left(\frac{4n_z\pi m}{N}\right)\big [(&&-1)^{n_y}W(1,m)\IEEEnonumber\\
&&+\>W(0,m)\big] . 
\label{eq:sq_app}
\IEEEeqnarraynumspace 
\end{IEEEeqnarray}
From the above Eq. (\ref{eq:sq_app}), it follows that the  structure factor at $(\pi,\pi)$ is given by
\begin{IEEEeqnarray*}{rCl}
S(\pi,\pi) &=& \sum_{m=1}^{N/2}(-1)^m[-W(1,m)+W(0,m)]\\
&=& \left(\sum_{m_{\text{even}}}-\sum_{m_{\text{odd}}}\right)
[-W(1,m)+W(0,m)] ,
\IEEEyesnumber
\IEEEeqnarraynumspace
\label{eq:app_spipidef}
\end{IEEEeqnarray*}
 at $(0,\pi)$ is given by
\begin{IEEEeqnarray*}{rCl}
S(0,\pi) &=& \sum_{m=1}^{N/2}(-1)^m[W(1,m)+W(0,m)]\\
&=& \left(\sum_{m_{\text{even}}}-\sum_{m_{\text{odd}}}\right)
[W(1,m)+W(0,m)] ,
\IEEEyesnumber
\IEEEeqnarraynumspace
\label{eq:app_s0pidef}
\end{IEEEeqnarray*}
 at $(\pi,0)$ is given by
\begin{IEEEeqnarray*}{rCl}
S(\pi,0) &=& \sum_{m=1}^{N/2}[-W(1,m)+W(0,m)]\\
&=& \left(\sum_{m_{\text{even}}}+\sum_{m_{\text{odd}}}\right)
[-W(1,m)+W(0,m)] ,
\IEEEyesnumber
\IEEEeqnarraynumspace
\label{eq:app_spi0def}
\end{IEEEeqnarray*}
and at $(0,0)$ is given by
\begin{IEEEeqnarray*}{rCl}
S(0,0) &=& \sum_{m=1}^{N/2}[W(1,m)+W(0,m)]\\
&=& \left(\sum_{m_{\text{even}}}+\sum_{m_{\text{odd}}}\right)
[W(1,m)+W(0,m)] .
\IEEEyesnumber
\IEEEeqnarraynumspace
\label{eq:app_s00def}
\end{IEEEeqnarray*}

Now, we calculate the various terms in the above equations as follows:
\begin{IEEEeqnarray*}{rCl}
&&\sum_{m_{\text{even}}}W(0,m)\\
&&\quad=
\frac{4}{N}\sum_{m_{\text{even}}}\sum_{j,k}\left[\langle n_{z^2;j,k}n_{z^2;j,k+m}\rangle-\frac{N_p^2}{N^2}\right]\\
&&\quad=\frac{4}{N}\sum_{j,k}\left[\langle n_{z^2;j,k}\sum_{m_{\text{even}}} n_{z^2;j,k+m}\rangle
-\sum_{m_{\text{even}}}\frac{N_p^2}{N^2}\right]\\
&&\quad=\frac{4}{N}\sum_{j}\left[\sum_{k_{\text{even}}}\langle n_{z^2;j,k}\sum_{m_{\text{even}}} n_{z^2;j,k+m}\rangle\right.\\
&&\qquad\left.+\sum_{k_{\text{odd}}}\langle n_{z^2;j,k}\sum_{m_{\text{even}}} n_{z^2;j,k+m}\rangle
-\sum_k\sum_{m_{\text{even}}}\frac{N_p^2}{N^2}\right]\\
&&\quad=\frac{4}{N}\sum_{j}\left[\sum_{k_{\text{even}}}\langle n_{z^2;j,k}\hat{N}_{j,e}\rangle
+\sum_{k_{\text{odd}}}\langle n_{z^2;j,k}\hat{N}_{j,o}\rangle
-\frac{N_p^2}{8}\right],
\end{IEEEeqnarray*}
where $\hat{N}_{j,e}=\sum_{k_{\text{even}}}n_{z^2;j,k}$ $\big (\hat{N}_{j,o}=\sum_{k_{\text{odd}}}n_{z^2;j,k}\big )$
is the number operator for the total number of particles occurring in the  even (odd) sublattice of the $j$-th chain.
Then, it follows that
\begin{IEEEeqnarray*}{rCl}
&& \!\!\!\! \sum_{m_{\text{even}}}W(0,m) \\
&& =
\frac{4}{N}\sum_{j}\left[\langle\hat{N}_{j,e} \hat{N}_{j,e}\rangle
+\langle \hat{N}_{j,o}\hat{N}_{j,o}\rangle
-\frac{N_p^2}{8}\right] \\
&& =
\frac{4}{N}\left[\langle\hat{N}_{1,e}^2 \rangle
+\langle \hat{N}_{1,o}^2\rangle
+\langle\hat{N}_{2,e}^2 \rangle
+\langle \hat{N}_{2,o}^2\rangle
-\frac{N_p^2}{4}\right] .
\IEEEyesnumber
\IEEEeqnarraynumspace
\label{eq:app_evenw0m}
\end{IEEEeqnarray*}
Next, we observe that
\begin{IEEEeqnarray*}{rCl}
&&\sum_{m_{\text{even}}}W(1,m)\\
&&\quad=
\frac{4}{N}\sum_{m_{\text{even}}}\sum_{j,k}\left[\langle n_{z^2;j,k}n_{z^2;j+1,k+m}\rangle-\frac{N_p^2}{N^2}\right]
\\
&&\quad=\frac{4}{N}\sum_{j,k}\left[
\langle n_{z^2;j,k}\sum_{m_{\text{even}}}
n_{z^2;j+1,k+m}\rangle-\sum_{m_{\text{even}}}\frac{N_p^2}{N^2}\right]\\
&&\quad=\frac{4}{N}\sum_{j}\left[
\sum_{k_{\text{even}}}\langle n_{z^2;j,k}\sum_{m_{\text{even}}} n_{z^2;j+1,k+m}\rangle\right.\\
&&\left.\qquad+\sum_{k_{\text{odd}}}\langle n_{z^2;j,k}\sum_{m_{\text{even}}}
n_{z^2;j+1,k+m}\rangle
-\sum_k\sum_{m_{\text{even}}}\frac{N_p^2}{N^2}\right]\\
&&\quad=\frac{4}{N}\sum_{j}\left[
\sum_{k_{\text{even}}}\langle n_{z^2;j,k}\hat{N}_{j+1,e}\rangle \right . \\
&& \left . \qquad \qquad \qquad +\sum_{k_{\text{odd}}}\langle n_{z^2;j,k}
\hat{N}_{j+1,o}\rangle-\frac{N_p^2}{8}
\right]\\
&&\quad=
\frac{4}{N}\sum_{j}\left[
\langle \hat{N}_{j,e}\hat{N}_{j+1,e}\rangle
+\langle\hat{N}_{j,o} \hat{N}_{j+1,o}\rangle
-\frac{N_p^2}{8}\right] \\
&&\quad=\frac{4}{N}\left[
2\langle \hat{N}_{1,e}\hat{N}_{2,e}\rangle
+2\langle\hat{N}_{1,o} \hat{N}_{2,o}\rangle
-\frac{N_p^2}{4}\right] .
\IEEEyesnumber
\label{eq:app_evenw1m}
\end{IEEEeqnarray*}
Similarly, we also obtain
\begin{IEEEeqnarray*}{rCl}
&&\!\!\!\!\!\! \sum_{m_{\text{odd}}}W(0,m)\\
&&\quad=
\frac{4}{N}\sum_{m_{\text{odd}}} 
\sum_{j,k}\left[\langle n_{z^2;j,k}n_{z^2;j,k+m}\rangle-\frac{N_p^2}{N^2}\right]\\
&&\quad=\frac{4}{N}\sum_{j,k}\left[
\langle n_{z^2;j,k}\sum_{m_{\text{odd}}} n_{z^2;j,k+m}\rangle
-\sum_{m_{\text{odd}}}\frac{N_p^2}{N^2}\right]\\
&&\quad=\frac{4}{N}\sum_{j}\left[\sum_{k_{\text{even}}}\langle n_{z^2;j,k}\sum_{m_{\text{odd}}} n_{z^2;j,k+m}\right.\rangle\\
&&\qquad\left.+\sum_{k_{\text{odd}}}\langle n_{z^2;j,k}\sum_{m_{\text{odd}}} n_{z^2;j,k+m}\rangle
-\sum_k\sum_{m_{\text{odd}}}\frac{N_p^2}{N^2}\right]\\
&&\quad=\frac{4}{N}\sum_{j}\left[\sum_{k_{\text{even}}}
\langle n_{z^2;j,k}\hat{N}_{j,o}\rangle 
 +\sum_{k_{\text{odd}}}\langle n_{z^2;j,k}\hat{N}_{j,e}\rangle
-\frac{N_p^2}{8}\right]\\
&&\quad=\frac{4}{N}\sum_{j}\left[\langle\hat{N}_{j,e} \hat{N}_{j,o}\rangle
+\langle \hat{N}_{j,o}\hat{N}_{j,e}\rangle
-\frac{N_p^2}{8}\right] \\
&&\quad=\frac{4}{N}\left[2\langle\hat{N}_{1,e} \hat{N}_{1,o}\rangle
+2\langle \hat{N}_{2,e}\hat{N}_{2,o}\rangle
-\frac{N_p^2}{4}\right].
\IEEEyesnumber
\label{eq:app_oddw0m}
\end{IEEEeqnarray*}
Lastly, we get
\begin{IEEEeqnarray*}{rCl}
&& \!\!\!\! \sum_{m_{\text{odd}}}W(1,m)\\
&&\quad=
\frac{4}{N}\sum_{m_{\text{odd}}}\sum_{j,k}\left[\langle n_{z^2;j,k}n_{z^2;j+1,k+m}\rangle-\frac{N_p^2}{N^2}\right]
\\
&&\quad=\frac{4}{N}\sum_{j,k}\left[
\langle n_{z^2;j,k}\sum_{m_{\text{odd}}}
n_{z^2;j+1,k+m}\rangle
-\sum_{m_{\text{odd}}}\frac{N_p^2}{N^2}\right]\\
&&\quad=\frac{4}{N}\sum_{j}\left[
\sum_{k_{\text{even}}}\langle n_{z^2;j,k}\sum_{m_{\text{odd}}} n_{z^2;j+1,k+m}\rangle\right.\\
&&\qquad\left.+\sum_{k_{\text{odd}}}\langle n_{z^2;j,k}\sum_{m_{\text{odd}}}
n_{z^2;j+1,k+m}\rangle
-\sum_k\sum_{m_{\text{odd}}}\frac{N_p^2}{N^2}\right]\\
&&\quad=\frac{4}{N}\sum_{j}\left[
\sum_{k_{\text{even}}}\langle n_{z^2;j,k}\hat{N}_{j+1,o}\rangle
\right . \\
&& \qquad \qquad \qquad \left .
+\sum_{k_{\text{odd}}}\langle n_{z^2;j,k}
\hat{N}_{j+1,e}\rangle
-\frac{N_p^2}{8}\right]\\
&&\quad=\frac{4}{N}\sum_{j}\left[
\langle \hat{N}_{j,e}\hat{N}_{j+1,o}\rangle
+\langle\hat{N}_{j,o} \hat{N}_{j+1,e}\rangle
-\frac{N_p^2}{8}\right] \\
&&\quad=\frac{4}{N}\left[
2\langle \hat{N}_{1,e}\hat{N}_{2,o}\rangle
+2\langle\hat{N}_{1,o} \hat{N}_{2,e}\rangle
-\frac{N_p^2}{4}\right].
\label{eq:app_oddw1m}
\IEEEyesnumber
\end{IEEEeqnarray*}
Using the expressions obtained
in 
 Eqs.~\eqref{eq:app_evenw0m}, \eqref{eq:app_evenw1m}, \eqref{eq:app_oddw0m}, and \eqref{eq:app_oddw1m}, 
we simplify Eqs.~\eqref{eq:app_spipidef}, \eqref{eq:app_s0pidef},\eqref{eq:app_spi0def}, and \eqref{eq:app_s00def} as follows:
\begin{IEEEeqnarray}{rCl}
S(\pi,\pi)= \frac{4}{N}\langle[(\hat{N}_{1,e}-\hat{N}_{1,o})+(\hat{N}_{2,o}-\hat{N}_{2,e})]^2\rangle,
\label{eq:app_spipi}
\IEEEeqnarraynumspace
\end{IEEEeqnarray}
\begin{IEEEeqnarray}{rCl}
S(0,\pi)= \frac{4}{N}\langle[(\hat{N}_{1,e}-\hat{N}_{1,o})+(\hat{N}_{2,e}-\hat{N}_{2,o})]^2\rangle, 
\label{eq:app_s0pi}
\IEEEeqnarraynumspace
\end{IEEEeqnarray}
\begin{IEEEeqnarray}{rCl}
S(\pi,0)= \frac{4}{N}\langle[(\hat{N}_{1,e}+\hat{N}_{1,o})-(\hat{N}_{2,e}+\hat{N}_{2,o})]^2\rangle,
\label{eq:app_spi0} 
\IEEEeqnarraynumspace
\end{IEEEeqnarray}
and
\begin{IEEEeqnarray}{rCl}
S(0,0)= \frac{4}{N}\langle[\hat{N}_p^2-N_p^2]\rangle,
\label{eq:app_s00} 
\IEEEeqnarraynumspace
\end{IEEEeqnarray}
where $\hat{N}_p \equiv (\hat{N}_{1,e}+\hat{N}_{1,o})+(\hat{N}_{2,e}+\hat{N}_{2,o})$ 
is the number operator for the total number of particles.
\begin{widetext}
\section{Evaluation of dynamic susceptibility for interacting CJT chains at weak coupling}
\label{app:weakcoupling}
In this appendix, we evaluate the real part of the dynamic susceptibility for interacting CJT chains in the weak-coupling regime.
A simple analytic expression for the  dynamic susceptibility is obtained as follows:
\begin{IEEEeqnarray*}{rCl}
\!\!\!\! {\rm Re}\>\chi_0(p,q;\omega_0)
&=&\frac{1}{N}\sum_{m\neq0} \Bigg[
\frac{|\langle m|_{el} \rho_{p,q}|0\rangle_{el}|^2}{\omega_0-\xi_{m0}}
-\frac{|\langle m|_{el} \rho_{-p,-q}|0\rangle_{el}|^2}{\omega_0+\xi_{m0}}
\Bigg] \\
&=&\frac{1}{N}\sum_{p^{\prime},q^{\prime}} \Bigg[
\frac{(1-n_{z^2;p^{\prime}+p,q^{\prime}+q})n_{z^2;p^{\prime},q^{\prime}}}
{\omega_0-(\epsilon_{p^{\prime}+p,q^{\prime}+q}-\epsilon_{p^{\prime},q^{\prime}})}
-\frac{(1-n_{z^2;p^{\prime}-p,q^{\prime}-q})n_{z^2;p^{\prime},q^{\prime}}}
{\omega_0+(\epsilon_{p^{\prime}-p,q^{\prime}-q}-\epsilon_{p^{\prime},q^{\prime}})}
\Bigg]\\
&=&\frac{1}{N}\sum_{p^{\prime},q^{\prime}} \Bigg[
\frac{n_{z^2;p^{\prime},q^{\prime}}}
{\omega_0-(\epsilon_{p^{\prime}+p,q^{\prime}+q}-\epsilon_{p^{\prime},q^{\prime}})} 
-\frac{n_{z^2;p^{\prime},q^{\prime}}}
{\omega_0+(\epsilon_{p^{\prime}+p,q^{\prime}+q}-\epsilon_{p^{\prime},q^{\prime}})}
\Bigg]\\
&=&\frac{1}{2\pi}\int^{k_F}_{-k_F}\Bigg[
\frac{dq^{\prime}}
{\omega_0-4t\sin(q^{\prime}+\frac{q}{2})\sin(\frac{q}{2})}
-\frac{dq^{\prime}}
{\omega_0+4t\sin(q^{\prime}+\frac{q}{2})\sin(\frac{q}{2})}
\Bigg]\\
&=&\frac{1}{2\pi}\int^{k_F}_{-k_F}
\frac{\gamma}{\omega_0}\Bigg[
\frac{dq^{\prime}}
{\gamma-\sin(q^{\prime}+\frac{q}{2}))}-\frac{dq^{\prime}}{\gamma+\sin(q^{\prime}+\frac{q}{2}))}
\Bigg] ,
\IEEEyesnumber
\end{IEEEeqnarray*}
where $\gamma=\frac{\omega_0}{4t\sin(\frac{q}{2})}$ and use has been made of the facts that
there is reflection symmetry and that  $\epsilon_{{p},{q}}$
is independent of the value of $p$ because there is no inter-chain transport.
Integrating the above equation for the case $\gamma<1$, we obtain 
\begin{IEEEeqnarray*}{rCl}
{\rm Re}\>\chi_0(p,q;\omega_0)
&=&\frac{1}{2\pi\omega_0}
\frac{\gamma}{\sqrt{1-\gamma^2}}
\Bigg\{\ln\Bigg[\frac{\big(1+\sqrt{1-\gamma^2}\big)^2-\big\{\gamma\tan\big(\frac{k_F}{2}+\frac{q}{4}\big)\big\}^2}
{\big(1-\sqrt{1-\gamma^2}\big)^2-\big\{\gamma\tan\big(\frac{k_F}{2}+\frac{q}{4}\big)\big\}^2}\Bigg]\\
&&\quad-\ln\Bigg[\frac{\big(1+\sqrt{1-\gamma^2}\big)^2-\big\{\gamma\tan\big(-\frac{k_F}{2}+\frac{q}{4}\big)\big\}^2}
{\big(1-\sqrt{1-\gamma^2}\big)^2-\big\{\gamma\tan\big(-\frac{k_F}{2}+\frac{q}{4}\big)\big\}^2}\Bigg]\Bigg\} .
\IEEEyesnumber
\end{IEEEeqnarray*}

\end{widetext}

\pagebreak
\section{Popular Summary}
Understanding the exotic phenomena (such as colossal magentoresistance,
coexisting charge-, spin-, and orbital-orderings) in bulk manganites 
and designing artificial structures  (such as heterostructures,  
quantum wires, and quantum dots) using manganites is of immense fundamental
interest  and also of huge technological importance (especially for
development of electronic and spintronic devices). 

To  study  emergent ordering, exploit device potential,
and guide material synthesis in these complex magnetic oxides,
effective Hamiltonians are needed for various interactions.
Except for the cooperative electron-phonon interaction (EPI),
effective Hamiltonians (that reasonably model the essential physics)
have been derived for all other interactions. For
instance, large Hund’s coupling can be mimicked by the double
exchange model, interactions between localized spins (that
have strong on-site repulsion) can be described by superexchange
model, and Hubbard Coulombic interaction can be modelled 
by dynamical mean-filed theory. The main challenges for modelling
cooperative EPI in manganites have been mathematical analysis of
the quantum-phonon effects and resolving the 
ambiguity in the nature of the electron-phonon interaction.
Here, using a controlled analytic treatment of quantum phonons,
we show that weak coupling and strong coupling produce different
experimentally observable effects.

Since manganites with C-type antiferromaganetic ordering can
be effectively considered as chains, we study the effect of 
electron-phonon coupling in interacting chains. At strong coupling, 
our theory predicts a conducting charge-density-wave (CDW) with 
charge residing predominantly in one sub-lattice; hence, the
ordering wavevector is independent of density. On the other hand,
at weak coupling, our theory shows that an insulating CDW
results with a wavevector that varies linearly with density. 
Thus, based on the experimental determination of the 
charge-ordering wavevectors (at various doping values) in
manganites that exhibit C-type antiferromagnetism, we
believe that our theory offers an opportunity for identifying
the regime of electron-phonon coupling.


\begin{thebibliography}{99}
\bibitem{gorsh_quest}
The interpretation of a CDW, with $q=0.25a^*$ for ${\rm La_{0.25}Ca_{0.75}MnO_3}$, reported in Ref. \onlinecite{nucara}
has been questioned by the authors of Refs. \onlinecite{gorshunov,gorshunov2}.
\end{thebibliography}
\end{document}